\documentstyle[aps,psfig,preprint,tighten,floats]{revtex}

\begin{document}

\title{Channel Coupling in $A(\vec{e},e^\prime \vec{N})B$ Reactions}
\author{James J. Kelly}
\address{ Department of Physics, University of Maryland, 
          College Park, MD 20742 }
\date{\today}
\maketitle

\begin{abstract}
The sensitivity of momentum distributions, recoil polarization observables,
and response functions for nucleon knockout by polarized
electron scattering to channel coupling in final-state interactions
is investigated using a model in which both the distorting and the
coupling potentials are constructed by folding density-dependent 
nucleon-nucleon effective interactions with nuclear transition densities.
Elastic reorientation, inelastic scattering, and charge exchange
are included for all possible couplings within the model space.
Calculations for $^{16}$O are presented for 200 and 433 MeV ejectile
energies, corresponding to proposed experiments at MAMI and TJNAF,
and for $^{12}$C at 70 and 270 MeV, corresponding to experiments
at NIKHEF and MIT-Bates.
The relative importance of charge exchange decreases as the ejectile
energy increases, but remains significant for 200 MeV.
Both proton and neutron knockout cross sections for large recoil momenta, 
$p_m > 300$ MeV/c, are substantially affected by inelastic couplings even
at 433 MeV.
Significant effects on the cross section for neutron knockout are also 
predicted at smaller recoil momenta, especially for low energies.
Many of the response functions and polarization observables for nucleon 
knockout are quite sensitive to the coupling scheme, especially those which 
vanish in the absence of final-state interactions.
Polarization transfer for proton knockout is insensitive to channel coupling,
even for fairly low ejectile energies, 
but polarization transfer for neutron knockout retains nonnegligible
sensitivity to channel coupling for energies up to about 200 MeV.
The present results suggest that possible medium modifications of neutron and 
proton electromagnetic form factors for $Q^2 \gtrsim 0.5$ (GeV/c)$^2$ can be 
studied using recoil polarization with relatively little uncertainty due to 
final-state interactions.
\end{abstract}

\section{Introduction}
\label{sec:intro}

Proton knockout by electron scattering has become established as
the most accurate method for measuring recoil momentum distributions
for nuclear single-hole states.
With the high resolution available at NIKHEF, precise measurements of 
distorted momentum distributions have been made for discrete states in many 
nuclei \cite{Dieperink90}.  
Recent reviews of nucleon electromagnetic knockout reactions can be found
in Refs.\ \cite{Kelly96,Boffi93,Boffi96}.
These studies have provided much information on the fragmentation of 
single-particle strength among various hole states in the residual nucleus.  
Although the missing momentum distributions for strongly excited states
generally agree quite well in shape with mean-field calculations,
the total strength observed is systematically lower than  
the single-particle strength.
The quenching of the single-particle strength is attributed to correlations
which spread that strength over broad ranges of both energy and momentum.
Therefore, evidence for these correlations has been sought in single-nucleon
knockout with large missing momentum for which correlations might be expected
to enhance the yield with respect to mean-field models.
However, because inelastic scattering and charge exchange contributions
to final-state interactions (FSI) can also enhance the yield for large 
missing momentum, it becomes important to extend the treatment of FSI
beyond the usual optical-model approach.

Another of the central problems of nuclear physics is to determine the
sensitivity of hadronic properties to the local baryonic density.
For example, an early hypothesis motivated by the EMC effect was that
the nucleon charge radius increases with density.
More recently, the quark-meson coupling (QMC) model has been used
to study the density-dependence of the nucleon electromagnetic form
factors \cite{Lu98a,Lu98b,Thomas98a} induced by coupling of their constituent 
quarks to the strong scalar and vector fields within nuclei.
However, because the effects predicted are relatively small at normal nuclear
densities, 
it will be very difficult to extract unambiguous results from measurements of 
cross sections for single-nucleon knockout from nuclei.
Fortunately, recoil polarization observables are expected to be much less 
vulnerable than cross sections to uncertainties in spectral functions, 
gauge ambiguities, and off-shell extrapolation of the single-nucleon current 
operator \cite{Kelly97}.
In the one-photon exchange approximation, the ratio between the longitudinal
and coplanar transverse polarization  transfers, $P^\prime_L / P^\prime_S$,
is proportional to the ratio between electric and magnetic form factors,
$G_E/G_M$, and this relationship is relatively insensitive to distortion
by the optical potential for the ejectile.
The primary objective of the present investigation is to determine the 
effect of channel coupling in final-state interactions, especially of
charge exchange, upon recoil polarization.

Ideally one should evaluate the nuclear electromagnetic current using a
many-body hamiltonian which accurately describes both bound and scattering 
states.
Calculations for $^{16}$O$(e,e^\prime N)$ have been performed 
for $T_N \approx 70$ -- 100 MeV by 
Ryckebusch {\it et al.} \cite{Ryckebusch89a}
using an HF-RPA model based upon a Skyrme interaction \cite{Ryckebusch88}.
The roles of channel coupling and two-body currents at large missing momentum
have also been investigated recently for $T_N \approx 100$ MeV
by van der Sluys {\it et al}.\ \cite{vdSluys96}.
Both bound and continuum wave functions are generated within the
Hartree-Fock (HF) mean field for the Skyrme interaction.
The current operator is also based upon the HF hamiltonian.
Thus, this approach preserves gauge invariance and avoids orthogonality 
defects.
On the other hand, because the mean field is real, attenuation of the 
scattered flux must be described by explicit coupling to the open channels.
Coupling to all single-nucleon emission channels is included within the 
random phase approximation (RPA), but more complicated configurations
are omitted.
Hence, although this model is internally consistent, its description 
of the final state interactions is limited to low ejectile energies
and is not suitable for the upcoming experiments at MAMI and TJNAF.

Jeschonnek {\it et al}.\ \cite{Jeschonnek94}
use a continuum RPA model in which coupling between single-nucleon
emission channels is treated microscopically while
coupling to more complicated channels is approximated using
a phenomenological optical model.
Coupling potentials were constructed using either a bound-state G-matrix 
based upon the Bonn potential \cite{Nakayama84} 
or the Franey-Love parametrization of the  t-matrix \cite{FL85}.
Although this approach provides a more realistic model of absorption,
it is now well-established that the nucleon-nucleon effective interaction
is strongly density-dependent and cannot be accurately represented by
a t-matrix; nor is the bound-state G-matrix appropriate for higher
ejectile energies.

The local density approximation (LDA) based upon density-dependent
empirical effective interactions (EEI) does provide accurate fits to
proton elastic, inelastic, and charge-exchange scattering for energies
above 100 MeV \cite{Kelly89b,Seifert93,Kelly91a,Flanders91}.
The density dependence of effective interactions constructed for infinite
nuclear matter, usually with $G$-matrix formalisms, is parametrized and
the parameters are adjusted to fit proton elastic and inelastic scattering 
data for self-conjugate targets using states whose transition densities are
measured by electron scattering.
Both the distorting and the scattering potentials are based upon the same
effective interaction, which is fitted using a self-consistency procedure.
Sensitivity to the density dependence of the effective interaction is
provided by use of both interior-peaked and surface-peaked transition 
densities.
It has been shown that the empirical effective interaction is essentially
independent of both state and target and that interactions fitted to
inelastic scattering data  provide good fits to elastic scattering whether
or not those data are included in the analysis 
\cite{Kelly91c,Seifert93}.
The EEI model also provides accurate predictions for proton absorption and
neutron total cross section data \cite{Kelly96b}.
Unlike optical-model analyses of elastic scattering data, which are sensitive
only to the asymptotic properties of the wave function, represented by
phase shifts, the overlap with interior-peaked transition densities gives 
the EEI analysis of inelastic scattering data sensitivity to the interior
wave function.
Finally, it has been shown that the EEI model accurately describes proton
scattering data for $^9$Be, where channel coupling within the rotational 
band plays an important role \cite{Kelly92}, even though the interactions 
were fitted to data for $A \geq 16$.
Therefore, we believe that the local density approximation based upon
empirical effective interactions should provide a superior description of
final state interactions (FSI) for $A(\vec{e},e^\prime \vec{N})B$ reactions
at energies above 100 MeV.

	We have developed a coupled-channels model for 
$A(\vec{e},e^\prime \vec{N})B$ reactions which includes elastic
reorientation, inelastic scattering, and charge exchange in final state 
interactions (FSI) based upon density-dependent effective interations within
the local density approximation.
A simpler version of the model was used recently to analyze coupled-channel 
effects upon the distorted momentum distributions for the 
$^{10}$B$(e,e^\prime p)^9$Be and $^{10}$B$(\gamma, p)^9$Be reactions 
\cite{deBever94}.
Coupling  within the $^9$Be rotational band was evaluated using
density-dependent nucleon-nucleon interactions folded with transition
densities fitted to electron and proton scattering measurements.
For missing momenta greater than 300 MeV/c quadrupole coupling is
found to enhance the momentum distributions for $(e,e^\prime p)$
in quasiperpendicular kinematics by factors up to 3-5 for various states;
even larger effects are predicted for $(\gamma,p)$.
The model has since been extended to include charge-exchange coupling
and to produce response functions and polarization observables.
Some preliminary results were shown in Refs.\ \cite{Kelly96,Kelly95}.

In this paper we investigate the sensitivity of momentum distributions,
response functions, and recoil polarization observables for
$(\vec{e},e^\prime \vec{N})$ reactions to both inelastic and charge
exchange couplings in final-state interactions, 
emphasizing upcoming experiments that include recoil polarization.
Experiments A1/2-93 at MAMI \cite{A1/2-93} and 89-033 at 
TJNAF \cite{TJNAF89-033} will look for 
modifications of the helicity-dependent recoil polarization in the
$^{16}$O$(\vec{e},e^\prime \vec{p})$ reaction for $T_p \approx 200$ MeV
and 433 MeV, respectively.
In addition, experiment 89-003 at TJNAF will measure large missing momenta
and will separate $R_{LT}$ for $^{16}$O$(e,e^\prime p)$ for 
$T_p \approx 433$ MeV.
Therefore, in this paper we investigate the effects of coupling between
valence single-hole states in the $^{16}$O$(\vec{e},e^\prime \vec{N})$
reaction.
Sec. \ref{sec:formalism} presents the coupled-channels formalism,
Sec. \ref{sec:FSI} gives further details of the coupling potentials, and  
Sec. \ref{sec:observables+rsfns} describes the observables and 
response functions for single-nucleon knockout.
Results for representative cases are presented in Sections \ref{sec:12C}
and \ref{sec:16O}.
Our conclusions are summarized in Section \ref{sec:conclusions}.

\section{Coupled-channels formalism for single-nucleon knockout}
\label{sec:formalism}

\subsection{Coupled equations for electron scattering}  
\label{sec:cc-electron}

Suppose that the $e+A$ electronuclear system is described by a hamiltonian
of the form
\begin{equation}
 {\cal H} = K_A + H_{e} + {\cal H}_{A} + V_{eA}
\end{equation}
where $K_A$ is the kinetic energy of the nucleus,
$H_{e} = -i \vec{\alpha} \cdot \vec{\nabla}_{e} + m_e \beta$ 
describes the motion of a free electron, 
${\cal H}_{A}$ describes the internal dynamics of the nuclear system,
and $V_{eA}$ is the interaction between them.
State vectors for the complete electronuclear system satisfy
eigenvalue equations of the form ${\cal H} \Psi = E \Psi$,
where $E$ is the total energy.
Let $\Psi^{(+)}_\alpha$ represent an electronuclear wave function that
contains incoming Coulomb-distorted electron waves in channel $\alpha$ and 
outgoing waves in all open channels.
The electronuclear wave function can be factored according to
\begin{equation}
\label{eq:Psi}
\Psi^{(+)}_\alpha({\bf r}_e,{\bf r}_A) =  
\int d^3p^\prime_{A\alpha} \; g({\bf p}_{A\alpha} - {\bf p}^\prime_{A\alpha}) 
  \exp{(i {\bf p}^\prime_{A\alpha} \cdot{\bf r}_A) }
  \sum_\beta \xi^{(+)}_{\alpha\beta}({\bf r}_{e}) \psi^{(-)}_{A\beta} 
\end{equation}
where ${\bf p}_{A\alpha}$ and ${\bf r}_A$ are the momentum and position 
of the nuclear center of mass and
$\xi^{(+)}_{\alpha\beta}({\bf r}_{e})$ represents the motion of the 
electron. 
It is convenient to normalize the target wave packet to unity at the
asymptotic momentum, such that $g(0) = 1$.
The state vectors, $\psi_{A\beta}$, for the nuclear subsystem satisfy 
eigenvalue equations of the form
${\cal H}_A \psi_{A\beta} = m_{A\beta} \psi_{A\beta}$
where $m_{A\beta}$ is the invariant mass of the nuclear system in
channel $\beta$.
The summation over state labels $\beta$ is interpreted as a sum over
discrete and integral over continuum states.
The bound states include both elastic and inelastic electron scattering,
while the continuum states include single-nucleon knockout, two-nucleon
emission, and more complicated reaction channels.

Rawitscher investigated the asymptotic behavior of coupled electronuclear 
wave functions using the method of steepest descent \cite{Rawitscher97}
and demonstrated that when the proper boundary conditions are applied
to the continuum states of the nuclear subsystem, 
only outgoing waves that satisfy energy conservation survive.
The wave function $\xi^{(+)}_{\alpha\beta}$ contains a Coulomb wave in
channel $\alpha$ and outgoing spherical waves in all open channels.
Label $\alpha$ usually refers to the ground state of the target nucleus, 
but later we will also require electronuclear 
wave functions containing a Coulomb-distorted electron wave impinging 
instead upon an excited state of the nuclear system; 
that excited state may be unbound and may contain one or more ejectiles.
Consequently, the nuclear system is described by a wave function 
$\psi^{(-)}_{A\beta}$ that satisfies incoming boundary conditions.
Suppose that channel $\beta$ refers to a proton ejectile plus a bound
state of the $B=A-1$ nucleus.
The wave function $\psi^{(-)}_{A\beta}$ would then contain a 
Coulomb-distorted nucleon wave in channel $\beta$ and incoming spherical 
waves in all open channels 
(including both $\beta$ and $\beta^\prime \neq \beta$) and would be related
by time reversal to the wave function $\psi^{(+)}_{A\beta}$ that describes
proton scattering by state $\beta$ of target $B$.
Of course, one does not normally have access to scattering by excited
states.

Introducing distorting potentials, $U_{e\beta}({\bf r}_{e})$,
and projecting out the nuclear state $\beta$, 
we obtain coupled equations for the electron wave function that have the 
form
\begin{equation}
\label{eq:electron-cc}
 \left( E_{e\beta} - H_{e} - U_{e\beta} \right) \xi_{\alpha\beta} = 
 {\sum_{\gamma}} \left( V_{e\beta \gamma} - 
U_{e\gamma} \delta_{\beta\gamma} \right) \xi_{\alpha\gamma}
\end{equation}
where $E_{e\beta} = E - K_{A\beta} - m_{A\beta}$
is the electron center-of-mass energy for channel $\beta$
and where the coupling potentials are
\begin{equation}
 V_{e \beta \gamma} = \langle \psi_{A\beta} | V_{eA} | 
\psi_{A\gamma} \rangle \; .
\end{equation}
The equations for bound and continuum states of the residual nuclear system
are formally identical, provided that the summation and boundary conditions
are interpreted properly.
For bound nuclear states, we could minimize the residual elastic terms on the 
right-hand side of Eq.\ (\ref{eq:electron-cc}) by choosing
\begin{equation}
 U_{e\beta}({\bf r}_{e}) \approx \langle \psi_{A\beta} | V_{eA} | 
\psi_{A\beta} \rangle \; ,
\end{equation}
but this choice may not converge for unbound states.
Nevertheless, recognizing that the dominant electron-nucleus interaction is 
due to the spherical part of the elastic Coulomb potential, 
one generally chooses $U_{e\beta}({\bf r}_{e})$ to be the Coulomb 
potential produced by the charge density of the nuclear ground state and
neglects the interaction between the electron and the ejectiles that might
be present in channel $\beta$.
Presumably the effects of more complicated residual elastic terms can be 
evaluated perturbatively, if necessary.
Note that these elastic terms also include magnetic contributions and recoil
corrections to the static Coulomb potential.
Therefore, we define electron distorted waves as the solutions to the 
homogeneous equations, such that
\begin{equation}
\label{eq:electron-distortion}
 \left( E_{e\alpha} - H_{e} - U_{e\alpha} \right) 
\zeta_{\alpha}({\bf r}_{e}) = 0 \; 
\end{equation}
where $U_{e\alpha}$ is approximated by the ground-state Coulomb potential.

The transition matrix for inelastic transitions between initial state
$\alpha$ and final state $\beta$ can now be expressed in the prior
representation as 
\begin{equation}
{\cal M}_{\beta\alpha} = \int d^3r_e d^3r_A
\langle \Psi^{(-)}_\beta({\bf r}_e,{\bf r}_A) | V_{eA} |
\exp{(i {\bf p}_{A\alpha} \cdot{\bf r}_A) }
\zeta^{(+)}_{\alpha}({\bf r}_e) \psi^{(-)}_{A\alpha}   \rangle  
\end{equation}
where 
\begin{equation}
\label{eq:Psi-}
\Psi^{(-)}_\beta({\bf r}_e,{\bf r}_A) =  
\int d^3p^\prime_{A\beta} \; g({\bf p}_{A\beta} - {\bf p}^\prime_{A\beta}) 
  \exp{(i {\bf p}^\prime_{A\beta} \cdot{\bf r}_A) }
\sum_\gamma \xi^{(-)}_{\beta\gamma}({\bf r}_{e}) \psi^{(-)}_{A\gamma} 
\end{equation} 
is a complete, fully coupled, wave function
containing outgoing Coulomb waves in channel $\beta$ and incoming
waves in all open channels.
Therefore, we obtain a matrix element of the general form
\begin{equation}
\label{eq:general-matrix-element}
{\cal M}_{\beta\alpha} = 
\int d^3p^\prime_{A\beta} \; g({\bf p}_{A\beta} - {\bf p}^\prime_{A\beta}) 
\int d^3r_e d^3r_A
\exp{(i ({\bf p}_{A\alpha}-{\bf p}^\prime_{A\beta}) \cdot{\bf r}_A) } 
\sum_\gamma 
{\cal V}_{\beta\gamma\alpha}({\bf r}_{e},{\bf r}_{A})
\end{equation}
where
\begin{equation}
\label{eq:Veff}
{\cal V}_{\beta\gamma\alpha}({\bf r}_{e},{\bf r}_{A}) = 
\langle \xi^{(-)}_{\beta\gamma}({\bf r}_{e}) \psi^{(-)}_{A\gamma} | V_{eA} |
\zeta^{(+)}_{\alpha}({\bf r}_e) \psi^{(-)}_{A\alpha}   \rangle  
\end{equation}
is an effective electron-scattering potential obtained by
integration over all internal coordinates of the nuclear system.
The transition matrix element contains the effects of channel coupling 
produced by both the nucleon-nucleus and the electron-nucleus interactions.
The summation over the index $\gamma$ includes nuclear states excited by
final-state interactions between the electron and the nuclear system,
but these dispersion corrections are subsequently neglected.
Channel coupling between nuclear states excited by nucleon-nucleus
final-state interactions will be developed in 
Section \ref{sec:coupled-equations}. 
 
The expression derived by Rawitscher, Eq. (4.7) of Ref. \cite{Rawitscher97},
for the inhomogeneous driving terms for electron scattering is very similar 
to our Eqs. (\ref{eq:general-matrix-element}-\ref{eq:Veff}).
The primary difference is that we include a target wave packet to
facilitate later use of the effective momentum approximation to develop a 
practical approximation.
Another superficial difference is that we employ the prior representation,
and hence have the coupled electron wave function on the left-hand side of 
the transition matrix element, 
whereas he uses the post representation in which the coupled 
electron wave function appears on the right-hand side of his Eq. (2.16).
Hence, dispersion corrections appear as final-state interactions here
and as initial-state interactions in Ref. \cite{Rawitscher97},
but these representations should be equivalent in the absence of subsequent 
approximations.

Electromagnetic coupling between low-lying bound states is often
described as dispersion corrections.
For example, Mercer \cite{Mercer77} evaluated dispersion corrections 
due to coupling to inelastic excitations of the target in the 
$A(e,e^\prime)A^\prime$ reaction and found those effects to be quite small.
The present formalism also includes coupling of the electron to more 
complicated states of the nuclear system, including knockout channels,
but evaluation of these subtle effects would be very difficult
computationally and is omitted.  
A qualitative discussion of some of these issues has been given by
Rawitscher \cite{Rawitscher97}.
In the present work, we study single-nucleon knockout under conditions 
where the distortion of the electron wave functions is relatively small, 
namely high energies and light targets.
Therefore, we will neglect channel coupling that could arise from the
electron-nucleus interaction and employ a simple approximation for the 
electron wave functions, namely the effective momentum approximation.
In the next several sections we outline the approximations used to
perform practical calculations for single-nucleon knockout under
conditions where channel-coupling by nuclear FSI can be important.

\subsection{Single-nucleon knockout}
\label{sec:SNK}

For the present application we consider only the one-body
part of the nuclear electromagnetic current.
Hence, we approximate the electromagnetic interaction by the 
single-nucleon contribution
\begin{equation}
\label{eq:VeA}
V_{eA} \approx e^2 \int d^3r_N \; \hat{j}_\mu({\bf r}_e)
\frac{ \exp{( i \omega r_{eN} )} }{ r_{eN} } \hat{J}^\mu({\bf r}_{NA})
= e^2  \int d^3r_N \int \frac{d^3 q^\prime}{(2\pi)^3} \; \hat{j}_\mu
\frac{ \exp{( -i {\bf q}^\prime \cdot {\bf r}_{eN} )} }{Q^{\prime 2}} 
\hat{J}^\mu
\end{equation}
where $\hat{j}_\mu$ and $\hat{J}^\mu$ are the electron and nucleon
current operators, $Q^{\prime 2} = q^{\prime 2} - \omega^2$ is the 
photon virtuality, 
${\bf r}_{eN} = {\bf r}_{e} - {\bf r}_{N}$ is the separation between
the electron and the ejectile, and
${\bf r}_{NA} = {\bf r}_{N} - {\bf r}_{A}$ is the ejectile position
relative to the barycentric system.
Substituting Eqs. (\ref{eq:Psi}) and (\ref{eq:VeA}) into
Eq. (\ref{eq:general-matrix-element}), we find 
\begin{equation}
  {\cal M}_{\beta\alpha} \approx \sum_{\gamma}  
  \int \frac{d^3q^\prime}{(2\pi)^3} \; \frac{4 \pi \alpha}{Q^{\prime 2}} 
  g^*({\bf q} - {\bf q}^\prime)
  {\cal J}^e_{\beta\gamma\alpha}(-{\bf q}^\prime) \cdot 
  {\cal J}^N_{\gamma\alpha}( {\bf q}^\prime )
\end{equation}  
where ${\bf q}={\bf p}_{A\beta}-{\bf p}_{A\alpha}$ is the asymptotic
momentum transfer.
The matrix elements of the electron and nuclear currents are contracted 
and the fine structure constant, $\alpha$, should not to be confused
with state labels.
The electron and nuclear current matrix elements are
\begin{mathletters}
\label{eq:currents}
\begin{eqnarray}
 {\cal J}^{e\mu}_{\beta\gamma\alpha}({\bf q}) &=& \int d^3r_{e} \;
        \exp{( i {\bf q} \cdot {\bf r}_{e} )} \langle
   \xi^{(-)}_{\beta\gamma}({\bf r}_{e})| \hat{j}^\mu({\bf r}_{e}) |
   \zeta_{\alpha}({\bf r}_{e}) \rangle \\
 {\cal J}^{N\mu}_{\gamma\alpha}({\bf q}) &=& \int d^3r_{NA} \;
     \exp{( i {\bf q} \cdot {\bf r}_{NA} )} \langle 
 \psi^{(-)}_{A\gamma}  | \hat{J}^\mu({\bf r}_{NA}) | 
 \psi^{(-)}_{A\alpha} \rangle \; .
\end{eqnarray}
\end{mathletters}
Recognizing that it will be more convenient to express the nucleon
distorted waves and overlap functions relative to the residual nucleus, $B$,
than to the barycentric system, we rescale the charge and current
density operators using
\begin{equation}
 \hat{J}^\mu({\bf r}_{NA}) = \left( \frac{m_A}{m_B} \right)^3 
 \hat{J}^\mu({\bf r}_{NB})
\end{equation}
where ${\bf r}_{NB} = {\bf r}_N - {\bf r}_B = (m_A/m_B) {\bf r}_{NA}$
is the ejectile position relative to the residual nucleus.
The nuclear current then becomes
\begin{equation}
 {\cal J}^{N\mu}_{\gamma\alpha}({\bf q}) = \int d^3r_{NB} \;
     \exp{( i {\bf q} \cdot {\bf r}_{NA} )} \langle 
 \psi^{(-)}_{A\gamma}  | \hat{J}^\mu({\bf r}_{NB}) | 
 \psi^{(-)}_{A\alpha} \rangle \; .
\end{equation}
The appearance of $r_{NA}$ in the exponential is a familiar recoil
correction ({\it e.g.}, see \cite{Boffi79}).

For light targets and relatively small ejectile momenta,
it may be necessary to include electromagnetic interactions in which the 
momentum is received by the residual nucleus while the observed nucleon 
is a spectator.
These contributions, often called exchange terms, can be included using
a straightforward extension of the results presented here, 
but are negligible for applications involving energetic nucleons.
Two-body currents are more complicated and will not be considered in
the present work.

\subsection{Electron current}
\label{sec:electron-current}
 
We expect the final-state interactions between the nucleon ejectile and the 
residual nucleus to dominate over multiple hard scattering of the electron; 
therefore, we neglect inelastic contributions to the electron distorted waves
and approximate $\xi_{\beta\gamma} \approx \zeta_\beta \delta_{\beta\gamma}$.
Furthermore, the principal effect of Coulomb distortion of the electron wave 
functions for high-energy beams and light targets can be described using the 
effective momentum approximation (EMA) \cite{Jin93b,vdSluys97} 
\begin{equation}
\zeta \approx  \frac{\bar{k}_e}{k_e}  
\exp{(i \bar{{\bf k}}_e \cdot {\bf r} ) u(\bar{{\bf k}}_e)} 
\end{equation}
where $u(\bar{{\bf k}}_e)$ is a free Dirac spinor with spin variables and
where the local momentum
\begin{equation}
\label{eq:keff} 
    \bar{{\bf k}}_e = {\bf k}_e + f_Z \frac{\alpha Z}{R_Z} {\bf \hat{k}}  
\end{equation}
is increased relative to the asymptotic wave number $k_e$ by the action of 
the Coulomb potential \cite{Schiff56}.
Here $f_Z = 1.5$ corresponds to the electrostatic potential at the center of 
a uniformly charged sphere of radius $R_Z$.
An improved version of the EMA proposed by Kim and Wright {\it et al.}\
\cite{Kim97} should allow the present formalism to be applied to 
heavier targets, 
but has not yet been implemented in the coupled-channels code.

Thus, the electron current is approximated by
\begin{equation}
{\cal J}^{e\mu}_{\beta\gamma\alpha}({\bf q}) \approx 
{\cal J}^{e\mu}_{\rm eff}({\bf q}_{\rm eff}) \delta({\bf q}-{\bf q}_{\rm eff})
\delta_{\beta\gamma}
\end{equation}
where
\begin{equation}
{\cal J}^{e\mu}_{\rm eff}({\bf q}_{\rm eff}) = 
  \frac{\bar{k}_i \bar{k}_f}{k_i k_f}
  \bar{u}(\bar{{\bf k}}_f) \gamma^\mu u(\bar{{\bf k}}_i) \; .
\end{equation}   
Finally, we assume that the wave packets vary sufficiently slowly with
momentum that we can replace $g({\bf q} - {\bf q}_{\rm eff})$ by unity;
in any case, the shape of the wave packet can be extracted from the
definition of the differential cross section.  
Therefore, the transition matrix element reduces in the effective momentum 
approximation to
\begin{equation}
  {\cal M}_{\beta\alpha} \approx  
  \frac{4 \pi \alpha}{Q^2_{\rm eff}} \; 
  {\cal J}^e_{\rm eff}(-{\bf q}_{\rm eff}) \cdot 
  {\cal J}^N_{\beta\alpha}( {\bf q}_{\rm eff} )
\end{equation}  
where $Q^2_{\rm eff}={\bf q}^2_{\rm eff} - \omega^2$ 
and where the nuclear current, 
given by Eq. (\ref{eq:currents}), includes channel coupling by the
nuclear final-state interactions.

\subsection{Coupled equations for nuclear FSI}
\label{sec:coupled-equations}

Suppose that the residual nucleon-nucleus system is described by
a hamiltonian of the form
\begin{equation}
 {\cal H}_{A} = m_N + K_{NB} + H_B + V_{NB} 
\end{equation}
where $K_{NB}$ is the kinetic energy operator for relative motion, 
$H_B$ is the internal hamiltonian of the residual nucleus,
and $V_{NB}$ is the potential energy for the nucleon-nucleus interaction 
and is real.
We also include the ejectile mass, $m_N$, but neglect its possible 
internal excitations.
The orthonormal state vectors of the residual nucleus satisfy 
eigenvalue equations of the form
\begin{equation}
 H_B \Phi_{\beta} = m_{B\beta} \Phi_{\beta} 
\end{equation}
where $m_{B\beta}$ is the invariant mass of the residual nucleus
in channel $\beta$.  
Recognizing that it is impractical to retain the complete set of
eigenstates for the nuclear system, 
it is useful to introduce the model-space projection
operators $P$ and $Q$, where $P$ selects the states of interest (the
model space) and $Q=1-P$ selects the rest (excluded space) such that
$P^2=P$, $Q^2=Q$, and $PQ=QP=0$.
One would normally limit the model space to a set of states that are
strongly populated by the direct reaction $A(e,e^\prime N)B$, 
here valence states reached by single-nucleon knockout,
plus other states of interest that can be reached by final-state 
interactions, here $B(N, N^\prime)B^\prime$.

Using standard manipulations (e.g. Ref.\ \cite{FeshbachII}),
it is straightforward to show that projected state vectors within 
the model space satisfy eigenvalue equations of the form
\begin{equation}
\left( m_{A\alpha} - H_{\rm eff} \right) P\psi^{(+)}_{\alpha} = 0 \;,
\end{equation}
where the effective hamiltonian for the model space
\begin{equation}
\label{eq:Heff}
  H_{\rm eff} = H_{PP} + H_{PQ} (m_{A\alpha}^{+} - H_{QQ})^{-1} H_{QP}
\end{equation}
includes the effects of the excluded space and where 
$m_{A\alpha}^{+} = m_{A\alpha} + i\delta$ includes a positive 
infinitesimal $\delta$ to ensure outgoing boundary conditions.
We use the customary notation $C {\cal H}_A D = H_{CD}$, 
where $C,D \in \{P,Q\}$.
The effective hamiltonian depends upon the model space selected and
is complex, energy dependent, nonlocal, and
far too complicated for practical applications.
Therefore, it is customary to approximate the effective model-space
hamiltonian by
\begin{equation}
\label{eq:Ueff}
 {\cal H}_{\rm eff} \approx  m_N + K_{NB} + H_B + U 
\end{equation}
where $U$ is a complex effective interaction.
Although the formalism applies equally well to nonlocal effective
interactions, we will employ local approximations in our applications.
For elastic channels $U$ is identified with the optical potential, 
whereas for inelastic channels $U$ becomes a transition potential.
Although one often employs phenomenological optical potentials fitted
to elastic scattering data, we prefer to use microscopic potentials for
both elastic and inelastic scattering obtained by folding density-dependent
effective interactions with nuclear transition densities.

Thus, we can expand the model-space wave functions according to
\begin{equation}
P\psi^{(+)}_{A\gamma} = 
\sum_\eta \chi^{(+)}_{\gamma \eta}({\bf r}_{NB}) \Phi_\eta
\end{equation}
where the $\Phi_\eta$ are state vectors of the residual nucleus and
$\chi^{(+)}_{\gamma \eta}({\bf r}_{NB})$ is the coupled-channels wave 
function for relative motion containing incoming waves in channel $\gamma$
and outgoing waves in all states within the model space.
Separating the dominant distorting potentials from the smaller coupling 
terms, we now find that the channel wave functions satisfy coupled equations
of the form
\begin{equation}
\left(m_{A\gamma} - m_{B\eta} - m_{N\eta} - K_{NB}  - 
U_\eta \right)\chi_{\gamma \eta} = 
{\sum_\kappa}^\prime U_{\eta\kappa} \chi_{\gamma\kappa} 
\end{equation}
where $U_{\eta\kappa}$ are the coupling potentials and where 
$U_\eta$ contains the central and spin-orbit components of the 
elastic potential for channel $\eta$.
One could include the complete elastic potential on the left, but it is
computationally more convenient to place the small nonspherical parts of 
the elastic potential (if any) on the right.
The primed summation indicates that any elastic terms included on the left 
are excluded from the right.

We have decided for the present applications to express the coupled
equations in the form of relativized Schr\"odinger equations.
Although there exists no rigorous justification for this procedure,
it is common in analyses of nucleon-nucleus scattering to employ
a prescription which replaces the center-of-mass kinetic energy and its
corresponding operator by
\begin{mathletters}
\begin{eqnarray}
m_{A\gamma} - m_{B\eta} - m_{N\eta} &\rightarrow& 
\frac{k_\eta^2}{2\mu_\eta} \\
K_{NB} &\rightarrow& 
\frac{- \nabla_{NB}^2}{2\mu_\eta} 
\end{eqnarray}
\end{mathletters}
where $k_\eta$ is the exact relativistic wave number in the $NB$ system
and $\mu_\eta$ is the relativistic reduced energy for channel $\eta$.
This procedure gives the correct de Broglie wave length and reproduces
the correct relativistic density of states \cite{Ray92}. 
The coupled equations are then expressed in coordinate space as
\begin{equation}
\label{eq:cc-FSI}
  (\nabla^2 + k_\eta^2 - 2\mu_\eta U_\eta) \chi_{\gamma \eta} = 
   2\mu_\eta {\sum_\kappa}^\prime U_{\eta \kappa} \chi_{\gamma \kappa} \; .
\end{equation}

Alternatively, one could describe nucleon-nucleus final-state interactions 
using the Dirac equation by means of the replacement
\begin{equation}
m_N + K_{NB} + U \rightarrow 
-i \vec{\alpha} \cdot \vec{\nabla}_{NB} +  \beta (m_N + S) + V
\end{equation}
where $S$ and $V$ are Dirac scalar and vector potentials;
additional Dirac potentials may be present also.
However, this approach requires a relativistic treatment of the
nuclear structure and the inelastic scattering potentials,
which is generally more difficult than the relativized Schr\"odinger
approach.
Although several Dirac coupled-channels calculations for proton-nucleus
scattering have been performed using coupling potentials based upon the
collective model \cite{Kurth94,Mishra90}, 
we are interested in charge-exchange and single-particle transitions
which require a more microscopic treatment of the coupling potentials.
Fortunately, it has been shown that nucleon-nucleon interactions
for the relativistic impulse approximation can be represented in terms 
of equivalent density-dependent effective interactions
suitable for use in the relativized Schr\"odinger formalism
\cite{Furnstahl93,Kelly94a}.
Furthermore, there exist empirical effective interactions fitted to
nucleon-nucleus elastic and inelastic scattering data over a
wide range of energies.
Therefore, we chose to employ the relativized Schr\"odinger approach,
which is computationally simpler than coupled Dirac equations,
with scattering potentials based upon the impulse approximation.

\subsection{Effective current operator}
\label{sec:effective-current}

It is also important to recognize that use of an effective hamiltonian should
be accompanied by renormalization of the current operator \cite{Boffi82a}.
The requirement that model-space matrix elements of the effective 
current operator, $\hat{J}^\mu_{\rm eff}$, reproduce those of the bare
current operator, $\hat{J}^\mu$, acting on complete wave functions is 
expressed by the condition
\begin{equation}
\langle \psi_{A\gamma} |  \hat{J}^\mu         | \psi_{A\alpha} \rangle =
\langle \psi_{A\gamma} |P \hat{J}_{\rm eff}^\mu P | \psi_{A\alpha} \rangle \;. 
\end{equation}
Thus, one obtains the formal expression
\begin{eqnarray}
  \hat{J}^\mu_{\rm eff} = \hat{J}^\mu_{PP}
 &+& \hat{J}^\mu_{PQ} (E^+ - H_{QQ})^{-1}H_{QP} 
   + H_{PQ}(E^+ + \omega - H_{QQ})^{-1}\hat{J}^\mu_{QP} \\ \nonumber
  &+& H_{PQ}(E^+ + \omega - H_{QQ})^{-1}\hat{J}^\mu(E^+ - H_{QQ})^{-1}H_{QP} 
  \; ,
\end{eqnarray}
where $E$ is the energy of the initial state and $E+\omega$ is the energy
of the final state.
This expression was obtained first by Boffi {\it et al}.\ \cite{Boffi82a},
who further assumed that $\hat{J}^\mu_{PQ}=\hat{J}^\mu_{QP}=0$.
An alternative expression for the effective current operator in terms of
the Green's function for the coupled equations has been given by
Rawitscher \cite{Rawitscher97}.
However, these expressions are extremely complicated and have never been
evaluated for realistic nuclear models. 
Hence, we conform with the universal and usually implicit practice of
assuming without proof that $\hat{J}_{\rm eff} \approx \hat{J}$.

Furthermore, in the spirit of the effective momentum approximation,
we replace momentum operators appearing in the nuclear current operator 
by their asymptotic values.
This approximation is consistent with the level of other approximations 
implicit in the replacement of the effective current with an off-shell
current operator based upon the free single-nucleon current.
Moreover, this procedure greatly simplifies the evaluation 
of the transition matrix elements,
with the electromagnetic vertex function reducing to a matrix acting
upon nucleon spin.
Although it would be straightforward to evaluate the momentum operators
completely, the inherent ambiguities in the choice of current operator 
\cite{Kelly97}
do not justify the computational cost.

Finally, note that by reducing the effective current operator to a 
two-dimensional matrix acting on nucleon spins, 
the effective momentum approximation permits the nucleon current
operator to be evaluated in the lab frame even though the distorted
wave calculations are performed in the barycentric frame.
Hence, the current matrix elements in Eq.\ (\ref{eq:currents}) and
the corresponding electromagnetic response tensors are both evaluated
in the lab frame.

We used the $\bar{\Gamma}_{\rm cc1}$ off-shell vertex function 
of de Forest \cite{deForest83} with 
nucleon form factors from the model 3 of
Gari and Kr\"umpelmann \cite{Gari92a,Gari92b}.
Current conservation was enforced at the one-body level by modifying the
longitudinal component of the current, which is equivalent to evaluating
the Feynman matrix element in the Coulomb gauge.
However, significant ambiguities persist in the off-shell behavior of the
nucleon electromagnetic vertex operator, which have been investigated by
many authors, e.g.\ \cite{deForest83,FM,Naus90,Song91,Chinn92}, 
without a clear resolution.
We studied the consequences of the these ambiguities for recoil
polarization in the $A(\vec{e},e^\prime \vec{N})$ reaction
under conditions of interest to experiments presently being performed
at MAMI and TJNAF \cite{Kelly97}.
Nevertheless, we expect that the qualitative changes relative to standard
optical model distortion that are produced by couplings to specific final 
states will be largely independent of these ambiguities.

\subsection{Nuclear current}
\label{sec:nucleon-current}

We now specialize to the case where the initial state contains the
ground state of the target, and denote the nuclear current for
excitation of state $\beta$ as 
${\cal J}^{N\mu}_{\beta} = {\cal J}^{N\mu}_{\beta 0}$ 
and the corresponding transition matrix element as
${\cal M}_{\beta} = {\cal M}_{\beta 0}$.
Given that the nuclear electromagnetic current operator has been approximated 
by a one-body operator, 
it is now appropriate to make a parentage expansion for the ground-state
of the target, such that
\begin{equation}
P \psi_0({\bf r}_{NB}) =
  \sum_{\lambda\nu} c_{\lambda \nu} 
  \phi_{\lambda \nu}( {\bf r}_{NB} ) \Phi_\lambda
\end{equation}  
where $c_{\lambda \nu}$ is a parentage coefficient (or pickup amplitude)
and $\phi_{\lambda \nu}( {\bf r}_{NB} )$ is an overlap function which 
describes the amplitude for removing a nucleon with single-particle quantum 
numbers $\nu$ at position ${\bf r}_{NB}$ relative to the core and
leaving the residual nucleus in state $\Phi_\lambda$.
The overlap function
\begin{equation}
 \phi_{\lambda \nu}( {\bf r} ) = R_{\lambda\nu}(r)
\sum_{m_\ell,m_j}
\left\langle 
\begin{array}{cc|c} 
  \ell_\nu  & \frac{1}{2} & j_\nu \\
  m_\ell & m_j - m_\ell  & m_j
\end{array}
\right\rangle
\left\langle
\begin{array}{cc|c} 
  j_\nu  & I_\lambda & I_0 \\
  m_j & m_\lambda  & m_0
\end{array}
\right\rangle
Y_{\ell_\nu m_\ell}(\hat{r}) \chi_{m_j-m_\ell}
\end{equation}
includes a radial function and the usual coupling of spherical harmonics
and nucleon spinors to the spin of the residual nucleus, $I_\lambda$, 
to produce the target spin, $I_0$.
The angle brackets denote Clebsch-Gordon coefficients. 
The parentage coefficient, $c_{\lambda \nu}$, requires two indices
whenever $I_0 \ne 0$. 
By extending the procedures outlined in Sec.\ \ref{sec:coupled-equations},
one could in principle develop a set of coupled equations governing
the overlap functions \cite{Pinkston65}.
However, one expects $\phi_{\lambda \nu}( {\bf r}_{NB} )$ to resemble a 
bound-state wave function in the potential generated by the residual 
nucleus.
Furthermore, analyses of $(e,e^\prime p)$ data produce phenomenological
overlap functions which are consistent in shape with single-particle 
wave functions based upon mean-field (Hartree-Fock) calculations.
Hence, we employ either Hartree-Fock wave functions or
Woods-Saxon wave functions fitted to $(e,e^\prime p)$ data.
More refined calculations in the future could employ overlap functions
projected from correlated wave functions.

Substituting the parentage expansion, the nuclear current now becomes
\begin{equation}
 {\cal J}_\beta^{N\mu}({\bf q}_{\rm eff}) \approx 
\sum_{\lambda\nu} c_{\lambda\nu}
\int d^3r_{NB} \;
  \exp{( i {\bf q}_{\rm eff} \cdot {\bf r}_{NA} )} \langle
  \chi^{(-)}_{\beta\lambda}({\bf r}_{NB}) | 
  \hat{J}_{\rm eff}^\mu( {\bf r}_{NB} ) | 
  \phi_{\lambda\nu}({\bf r}_{NB}) \rangle  \; .
\end{equation}
Finally, using the effective momentum approximation for the nucleon current
operator, we obtain
\begin{equation}
 {\cal J}_\beta^{N\mu}({\bf q}_{\rm eff}) \approx 
\sum_{\lambda\nu} c_{\lambda\nu}
\int d^3r_{NB} \;
  \exp{( i {\bf q}_{\rm eff} \cdot {\bf r}_{NA} )} \langle
  \chi^{(-)}_{\beta\lambda}({\bf r}_{NB}) | 
  \hat{J}_{\rm eff}^\mu( {\bf p}_{m,{\rm eff}}+{\bf q}_{\rm eff}, 
                         {\bf p}_{m,{\rm eff}} ) | 
  \phi_{\lambda\nu}({\bf r}_{NB}) \rangle  
\end{equation}
where ${\bf p}_{m,{\rm eff}} = {\bf p}_{N\beta} - {\bf q}_{\rm eff}$ is the 
missing momentum determined by the ejectile momentum, 
${\bf p}_{N\beta}$, and the effective momentum transfer,
${\bf q}_{\rm eff}$.
Thus, the nucleon current operator,
$ \hat{J}_{\rm eff}^\mu( {\bf p}_{m,{\rm eff}}+{\bf q}_{\rm eff}, 
{\bf p}_{m,{\rm eff}} ) $,
has been reduced to a matrix that acts upon nucleon spin.

This is the central result of the effective-momentum approximation to the 
coupled-channels formalism for FSI in nucleon knockout by electron scattering.
A similar electroexcitation amplitude was proposed by Blok and
van der Steenhoven \cite{Blok87b} based upon more qualitative arguments
that exploit the similarity between knock-out and pick-up reactions.
The primary difference between this expression and the standard
distorted-wave approximation (DWA) is that the coupled-channels 
wave function replaces the usual distorted wave.
Thus, we recover the DWA by neglecting the FSI coupling potentials,
such that 
$\chi_{\beta\lambda} \rightarrow 
\chi^{(0)}_{\beta\lambda} \delta_{\beta\lambda}$.
Also note that for single-nucleon knockout from a spinless target,
one generally assumes that the overlap function is well approximated
by a unique single-particle wave function, such that
$c_{\lambda\nu} \rightarrow \sqrt{S_\lambda} \delta_{\lambda\nu}$ 
where $S_\lambda$ is the spectroscopic factor.
However, the coupled-channels approach is much more general.
For example, states for which the single-nucleon parentage coefficients
are vanishingly small can still be populated by final-state interactions
following an intermediate step that involves a state
that is strongly excited by single-nucleon knockout  
\cite{Blok87b,Kelly96,Rawitscher97}.

\section{Final-state interactions}
\label{sec:FSI}

\subsection{Partial-wave potentials}
\label{sec:pw-potentials}

For each pair of channels (with suppressed labels),  the 
scattering operator can be decomposed into a sum of products
\begin{equation}
  U = \sum_{\kappa \lambda} \nu_{\kappa \lambda}(r) {\cal P}_{\kappa \lambda} 
                                              \cdot {\cal T}_{\kappa \lambda}
\end{equation}
in which $\nu_{\kappa \lambda}(r)$ contains the dependence upon relative 
separation and where the multipole operators ${\cal P}_{\kappa \lambda}$ 
and ${\cal T}_{\kappa \lambda}$ depend only on the angular momenta and 
internal variables of the projectile and target, respectively.
The angular momentum transfer is designated $\lambda$, whereas $\kappa$ is
used to distinguish between different operators with the same multipolarity.
Partial wave potentials of the form
\begin{equation}
  U_{\beta \gamma J} = \langle l_\beta  s_\beta  j_\beta  I_\beta  J M |U|
                               l_\gamma s_\gamma j_\gamma I_\gamma J M \rangle
\end{equation}
then become
\begin{equation}
 U_{\beta \gamma J} = \sum_{\kappa \lambda} 
     \Gamma_{\lambda J}(j_\beta I_\beta; j_\gamma I_\gamma)
     \nu_{\beta \gamma \kappa \lambda}
\end{equation}
where the recoupling coefficient is
\begin{equation}
  \Gamma_{\lambda J}(j_\beta I_\beta; j_\gamma I_\gamma) =
   (-)^{j_\gamma + I_\beta+J} \hat{\jmath_\beta} \hat{I_\beta}
  \left\{ 
\begin{array}{lll} 
  j_\beta  & j_\gamma & \lambda \\
  I_\gamma & I_\beta  & J
\end{array}
\right\}
\end{equation}
and where
\begin{equation}
   \nu_{\beta \gamma \kappa \lambda}(r) = \nu_{\kappa \lambda}(r) 
   \langle l_\beta  s_\beta  j_\beta || {\cal P}_{\kappa \lambda} ||
           l_\gamma s_\gamma j_\gamma  \rangle
 \langle \beta I_\beta || {\cal T}_{\kappa \lambda} || \gamma I_\gamma \rangle
\end{equation}
are the appropriate multipole potentials, including angular and target
matrix elements.  
The orbital angular momentum $l_\beta$ is combined with the projectile
spin $s_\beta$ to give ${\bf j}_\beta = {\bf l}_\beta + {\bf s}_\beta$,
which is then coupled to the target spin $I_\beta$ to give the channel spin
${\bf J} = {\bf I}_\beta + {\bf j}_\beta$. 
Also note that $\hat{\jmath} = \sqrt{2j+1}$ for angular momenta.
A standard partial wave analysis of the coupled equations is made and
the equations for each channel spin are solved by an iterative technique
based upon that of Raynal \cite{Raynal72}.

The coupling potentials that would emerge from 
Eqs. (\ref{eq:Heff}-\ref{eq:Ueff})
depend upon the chosen model space and are complex, nonlocal, 
energy-dependent, and otherwise intractable for practical applications.
However, when the model space is a very small fraction of the available
phase space, the dependence of the effective hamiltonian upon the selection
of states should be negligible. 
Furthermore, for energetic nucleons and low-lying nuclear excitations, it
is reasonable to approximate the coupling potentials using the impulse
approximation based upon a density-dependent nucleon-nucleon interaction
that provides good descriptions of nucleon elastic, inelastic, and
charge-exchange scattering to similar states.
Therefore, we constructed both optical and coupling potentials by folding 
a local density-dependent nucleon-nucleon interaction with nuclear 
transition densities that describe the relevant aspects of the target 
structure.
Details of the implementation of the folding model may be found in
Ref.\ \cite{LEA}.

It has been shown that the isoscalar spin-independent central, $t^C_{00}$, 
and isoscalar spin-orbit, $t^{LS}_0$, components of the effective 
interaction depend strongly upon local density \cite{Kelly80}, 
but are essentially independent of target \cite{Seifert93}.
However, although nuclear matter theory provides a good qualitative
description of these effects, theoretical interactions are not yet
sufficiently accurate \cite{Kelly89a,Kelly89b}.  
Therefore, for energies above 100 MeV we employ the empirical effective 
interactions fitted to proton elastic and inelastic scattering data that are 
tabulated in Ref.\ \cite{Kelly94a}, performing interpolations with respect
to energy when needed.
For lower energies we use the density-dependent Paris-Melbourne effective 
interaction \cite{Dortmans94,Karataglidis95}.
All components of the effective interaction except tensor exchange were
included in the coupled-channels calculations.
The isoscalar components of coupling potentials include
the Cheon rearrangement factor \cite{Cheon85a,Cheon85b},
which has been shown to be essential to the consistency between elastic and 
inelastic scattering in the analysis of the empirical effective interaction 
\cite{Kelly89b}. 
Rearrangement corrections for isovector interactions are more complicated
but less important \cite{Cheon92} and are omitted.

\subsection{Model space}
\label{sec:model-space}

Each state in the model space can be populated by direct single-nucleon
knockout or by final-state interactions following excitation of another
member of the model space.
All possible couplings between members of the model space are included.
For a model space with $n$ states there will be $n(n+1)/2$ couplings
between states, each with several possible multipolarities depending
upon the spins involved.
For each multipolarity, there will be several potentials based upon 
various components of the nucleon-nucleon effective interaction. 

It is useful to distinguish between four types of coupling mechanisms.
Coupling potentials which do not change the state of the residual nucleus
but which are omitted from the distorting (optical) potentials are
classified as {\it elastic reorientation}; 
reorientations effects are often dominated by quadrupole potentials,
when possible, but also include other allowed multipolarities.
{\it Inelastic excitations} change the state of the residual nucleus
without changing its charge; because the residual nucleus often has
nonzero spin, several multipolarities are usually possible.
{\it Analog transitions} change the charge of the residual nucleus
without changing its spin, and also include several multipolarities 
when the spin is greater than zero.
Finally, {\it nonanalog charge-exchange} transitions change both the 
internal state and the charge of the residual nucleus.

The present formalism is sufficiently general to accommodate 
sophisticated structure models in which correlations spread the
single-particle strength over many fragments and modify the radial
overlap functions.
Such correlations would also affect the OBDME used to construct
coupling potentials between members of the model space.
However, because the computational cost increases rapidly with the
size of the model space, it is necessary to limit the model space to
the states of interest and those to which coupling is strongest. 

In the present paper we consider coupling between low-lying discrete
states with strong direct knockout amplitudes,
for which the most important effects of channel coupling are
likely to be dominated by the strongest fragments.
In an earlier paper \cite{Kelly96} we had studied indirect excitation of 
states with negligible direct knockout amplitudes and demonstrated that under
some conditions multistep processes dominate, thereby improving upon
the two-step calculations of Blok and van der Steenhoven \cite{Blok87b}.
However, we have also demonstrated that states of this type are excited 
too weakly to affect states with strong direct amplitudes and may be 
safely omitted from the model space.
Furthermore, we assume that the coupling of low-lying discrete states
to continuum states of the residual nucleus is adequately represented 
through continuum contributions to the imaginary parts of the optical and 
coupling potentials and that continuum states need not be included 
explicitly in the model space.
Finally, we assume for these exploratory calculations that the 
independent-particle model (IPM) provides an adequate representation
of single-nucleon knockout summed over related fragments.
Therefore, the calculations were performed using IPM parentage 
coefficients and comparisons to experimental data include spectroscopic 
factors to normalize the strengths of the observed fragments.

The IPM model space for $^{12}$C$(e,e^\prime N)$ consists of the 
$(1s_{1/2})^{-1}$ and $(1p_{3/2})^{-1}$ proton-hole states in $^{11}$B
reached by the $(e,e^\prime p)$ reaction and the analog states in
$^{11}$C reached by the $(e,e^\prime n)$ reaction, for a total of 4 states.
In addition to these $(1s_{1/2})^{-1}$ and $(1p_{3/2})^{-1}$ hole
states, the IPM model space for $^{16}$O$(e,e^\prime N)$ also includes 
the $(1p_{1/2})^{-1}$ hole states in $^{15}$N and $^{15}$O, 
for a total of 6 states.
In the context of the IPM model space, we speak of the $(1s_{1/2})^{-1}$ 
hole configuration as a discrete state even though its spreading width
is actually appreciable.
The underlying continuum of two-nucleon and multinucleon knockout
states then constitute the excluded space whose effects upon the reaction
would, in principle, be represented through their influence upon the 
effective hamiltonian and effective current operators. 
However, in practice simple approximations to these effective operators 
are employed as described above.

The parentage coefficients for pure hole states are given by  
$c_{\beta\gamma}=\sqrt{2j_\gamma+1}$ where $j_\gamma=I_\beta$
is the spin of the residual nucleus.
The OBDME for coupling between pure hole states are given by
\begin{equation}
\label{eq:hole-coupling}
  \langle \beta^{-1} || [a^\dagger_p \otimes a_h]_J || \gamma^{-1} \rangle
    = \hat{\jmath}_p \delta_{J,0} \delta_{ph} \delta_{\beta \gamma}
        + (-)^{j_\gamma + j_\beta - J} \frac{\hat{J}}{\hat{\jmath}_\beta} 
            \delta_{p \gamma} \delta_{h \beta}
\end{equation}
where the initial state is described as a hole in orbital $\gamma$
and the final state as a hole in orbital $\beta$ for an otherwise
closed-shell nucleus.
We generally assume that the $(J=0,T=0)$ term is already included in the
spherical optical potential, so that only the second term of 
Eq. (\ref{eq:hole-coupling}) contributes to the coupling potentials.

Overlap functions were represented by Woods-Saxon single-particle wave
functions and fitted to $(e,e^\prime p)$ data where available.
Very similar results are obtained using Hartree-Fock wave functions.
Possible modifications of the radial wave functions by short-range
correlations can be incorporated easily. 


\section{Observables and response functions for 
            $A(\vec{\lowercase{e}},\lowercase{e}^\prime \vec{N})B$}
\label{sec:observables+rsfns}

\subsection{Observables}
\label{sec:observables}

Nucleon knockout reactions of the type $A(\vec{e},e^\prime \vec{N})B$
initiated by a longitudinally polarized electron beam and for which the 
ejectile polarization is detected may be described by a doubly differential 
cross section of the form \cite{Giusti89,Picklesimer89}
\begin{equation}
\label{eq:recoil-polarization}
 \frac{d^5 \sigma _{hs}}{d\varepsilon _{f}d\Omega _{e}d\Omega _{N}} =
   \sigma _{0} \frac{1}{2} \left[1 + \bbox{P}\cdot \bbox{\sigma} 
   +h (A + \bbox{P}^{\prime} \cdot \bbox{\sigma})\right] 
\end{equation}
where $\varepsilon_i$ ($\varepsilon_f$) is the initial (final) electron
energy, $\sigma _{0}$ is the unpolarized cross section, $h$ is the
electron helicity, $s$ indicates the nucleon spin projection upon 
$\bbox{\sigma}$, $\bbox{P}$ is the induced polarization, $A$ is the electron
analyzing power, and $\bbox{P}^{\prime}$ is the polarization transfer 
coefficient.
Thus, the net polarization of the recoil nucleon $\bbox{\Pi}$ has two 
contributions of the form
\begin{equation}
   \bbox{\Pi} = \bbox{P} + h \bbox{P}^\prime
\end{equation}
where $\mid h \mid \leq 1$ is interpreted as the longitudinal beam 
polarization.

The recoil polarization is usually calculated with respect to a  
helicity basis in the barycentric frame defined by the basis vectors
\begin{mathletters}
\begin{eqnarray} 
\label{eq:lispnuc}
\bbox{\hat{L}} &=& \frac{\bbox{p}_N}{| \bbox{p}_N| } \\
\bbox{\hat{N}} &=& \frac{\bbox{q} \otimes \bbox{\hat{L}}}{ |\bbox{q} 
\otimes \bbox{\hat{L}}|} \\
\bbox{\hat{S}} &=& \bbox{\hat{N}} \otimes \bbox{\hat{L}} \; .
\end{eqnarray}
\end{mathletters}
However, since this basis is not well defined when $\bbox{q}$ and 
$\bbox{p}_N$ are either parallel or antiparallel,
these cases are conventionally handled by first rotating the reaction 
plane to $\phi_N$  as it would be in non-parallel kinematics,
and then taking the limit $\theta_{pq} \rightarrow 0^\circ$ 
or $\theta_{pq} \rightarrow 180^\circ$ as required.
Note that since the basis vectors $\bbox{\hat{S}}$ and $\bbox{\hat{N}}$ 
reverse directions when $\phi \rightarrow \phi+\pi$, 
the corresponding components of the recoil polarizations also tend to 
reverse sign even when there is no physical asymmetry with respect to $\phi$; 
this behavior is simply an artifact of the basis.

Alternatively, since the recoil polarization is measured in the laboratory 
frame, it is useful to employ a polarimeter basis in which
\begin{mathletters}
\begin{eqnarray} 
\label{eq:labpnuc}
\bbox{\hat{y}} &=& \frac{ \bbox{k}_i \otimes \bbox{k}_f }
                        {|\bbox{k}_i \otimes \bbox{k}_f|} \\
\bbox{\hat{x}} &=& \frac{ \bbox{\hat{y}} \otimes \bbox{p}_N }
                        {|\bbox{\hat{y}} \otimes \bbox{p}_N|} \\
\bbox{\hat{z}} &=& \bbox{\hat{x}} \otimes \bbox{\hat{y}} \; .
\end{eqnarray}
\end{mathletters}
One advantage of presenting the recoil polarization in the lab or
polarimeter basis, is that the recoil polarization components are continuous
as ${\bf p}_N$ moves through ${\bf q}$ from one side to the other.
Unlike $\bbox{\hat{S}}$ and $\bbox{\hat{N}}$, $\bbox{\hat{x}}$ and 
$\bbox{\hat{y}}$ do not reverse directions when $\phi \rightarrow \phi+\pi$.
For coplanar quasiperpendicular kinematics with $\bbox{\hat{y}}$ upwards,  
it has become conventional to assign positive missing momentum to ejectile 
momenta on the large-angle side of ${\bf q}$, such that $\phi=\pi$ and 
$\theta_{pq} > 0$.

The distorted missing momentum distribution $\rho^D({\bf p}_m,{\bf p}^\prime)$,
which is more properly called the reduced cross section,
is obtained by dividing the unpolarized differential cross section 
$\sigma_0$ by the elementary electron-nucleon
cross section $\sigma_{eN}$ 
for initial (final) nucleon momenta ${\bf p}_m$ (${\bf p}^\prime$), such that
\begin{equation}
 \rho^D({\bf p}_m,{\bf p}^\prime) =  \frac{\sigma_0}{K \sigma_{eN}}
\end{equation}
where
\begin{equation}
\sigma_{eN} = \frac{\varepsilon_f}{\varepsilon_i} \frac{\alpha^2}{Q^4}
\eta_{\mu\nu} {\cal W}^{\mu\nu}_{eN}
\end{equation}
is based upon the PWIA response tensor for off-shell kinematics and
does not include the phase-space factor $K$.
To be consistent, the $eN$ response tensor must be computed from
the same current operator and gauge used to evaluate the 
nuclear response tensor.
The normalization is determined by the requirement that in the plane-wave 
approximation the momentum distribution, $\rho_j(p_m)$, for a fully occupied 
orbital with total angular momentum $j$ be normalized to its occupancy, 
such that
\begin{equation}
 4\pi \int dp_m \; p_m^2 \; \rho_j(p_m) = 2j+1
\end{equation}
for the independent-particle shell model.

\subsection{Response functions}
\label{sec:rsfns}

Additional insight into the reaction mechanism can be obtained by
examining response functions.
In the one photon-exchange approximation  
the observables may be represented in terms of sums of products between 
kinematical factors which depend only on electron scattering kinematics 
and response functions which represent the dynamical content of the reaction.
The details of the response-function decomposition have been given
many times before and will be omitted here --- 
we employ the definitions and notation of Ref.\ \cite{Kelly96}.

However, distortion of the electron wave function perturbs the relationship
between the asymptotic electron-scattering kinematics and the momentum
transfer delivered by the hard virtual photon,
thereby introducing additional dependencies upon azimuthal angle, $\phi$, and
upon electron scattering kinematics.
Nevertheless, these effects are small enough for high-energy electrons and 
light targets to usefully employ the response function decomposition.
For our purposes it will be instructive and sufficient to display 
response functions obtained by neglecting electron distortion.

It is useful to distinguish between Class I response functions that would
remain finite in the absence of final-state interactions and 
Class II response functions which would vanish if FSI could be eliminated.
Clearly one expects Class II response functions to be more sensitive to
the details of final-state interactions than Class I.
Class I includes the unpolarized $R_L$, $R_T$, $R_{LT}$,
and $R_{TT}$ response functions, $R^{\prime N}_{LT}$, and 
both $R^{\prime m}_{LT}$ and $R^{\prime m}_{TT}$ with $m \in \{L,S\}$.
Class II includes $R^\prime_{LT}$, $R^N_L$, $R^N_T$, 
$R^N_{LT}$, $R^N_{TT}$, and both $R^m_{LT}$ and $R^m_{TT}$ with 
$m \in \{L,S\}$.

\subsection{Kinematics}
\label{sec:kinematics}

The invariant mass of the final nuclear system is given by
$W^2 = m_A^2 + 2m_A \omega - Q^2$.
For the purposes of describing the final-state interactions, it is
useful to define $T_0$ to be the ejectile energy in the rest frame of
the residual nucleus, such that
$W^2 = (m_N+m_B)^2 + 2m_B T_0$.
The value quoted for $T_0$ is evaluated for the ground-state of the nucleus
reached in the $(e,e^\prime p)$ reaction. 
Solving for $T_0$, we obtain
\begin{equation}
T_0 =\frac{m_A}{m_B} \left[ \omega - E_m - \frac{Q^2 + E_m^2}{2m_A} \right]
\end{equation}
where $E_m = m_N + m_B - m_A$ is the missing energy.
Similarly, the missing momentum is defined by 
${\bf p}_m = {\bf p}_N - {\bf q}$.

For each missing momentum distribution, we hold $W$, or equivalently $T_0$,
constant so that FSI can be computed for a unique total energy.
To minimize variations in electron distortion, the beam energy is also fixed.
Parallel kinematics are defined by the subsidiary condition $\theta_{pq}=0$
and $p_m=p_N-q$ is varied by adjusting both $\omega$ and ${\bf q}$ as required
to maintain both $\theta_{pq}=0$ and constant $T_0$.
Quasiperpendicular kinematics maintain constant $(\omega,{\bf q})$ and require
$p_m=0$ for the $(e,e^\prime p)$ ground-state transition when $\theta_{pq}=0$.
The missing momentum is varied by changing $\theta_{pq}$ and is conventionally
defined as positive when the ejectile emerges on the large-angle side of the
momentum transfer vector, such that $\theta_p > \theta_q$.
Hence, positive $p_m$ for quasiperpendicular kinematics corresponds to
an angle $\phi=180^\circ$ between the reaction and scattering planes.

\section{Channel coupling in $^{12}$C$(\lowercase{e,e}^\prime N)$ }
\label{sec:12C}

\subsection{Charge exchange for $T_0 = 70$ MeV}
\label{sec:12C-cx}

The role of charge exchange in neutron electromagnetic knockout
was first investigated by 
van der Steenhoven {\it et al.} \cite{vdSteenhoven87a}  
using the Lane model.
They predicted that the charge exchange contribution to $(e,e^\prime n)$
substantially increases the longitudinal response for that
predominantly transverse reaction. 
For example, their calculations for $^{12}$C$(e,e^\prime n)$ in parallel
kinematics give as much as an order of magnitude enhancement of 
the neutron missing momentum distribution.
However, Giusti and Pacati \cite{Giusti89} found only very small effects
using a similar model.
On the other hand, using a continuum RPA model,  
Jeschonnek {\it et al}.\ \cite{Jeschonnek94}
obtained intermediate results which show much larger charge exchange
contributions than Giusti and Pacati that remain considerably smaller than 
those of van der Steenhoven {\it et al}.
Their continuum RPA included coupling between states reached by both
p-shell and s-shell knockout and employed a more complete model of
final-state interactions that included spin-isospin components of the
effective interaction.

We performed similar calculations for $^{12}$C$(e,e^\prime N)$ using
kinematics based upon the NIKHEF conditions.
The electron beam energy was taken to be 461 MeV and
all calculations maintain a constant total energy in the final state
that is equivalent to a proton with 70 MeV kinetic energy incident
upon the ground state of $^{11}$B at rest.
For simplicity we approximate the ground state using the 
independent-particle model, such that the model space consists of the
four single-hole states reached by single-nucleon knockout.
Coupling between these states is described by transition potentials
obtained by folding the density-dependent Paris-Melbourne 
effective interaction for 65 MeV \cite{Dortmans94,Karataglidis95}
with single-particle transition densities as described above.  

\begin{figure}[htb] 
\centerline{ \strut\psfig{file=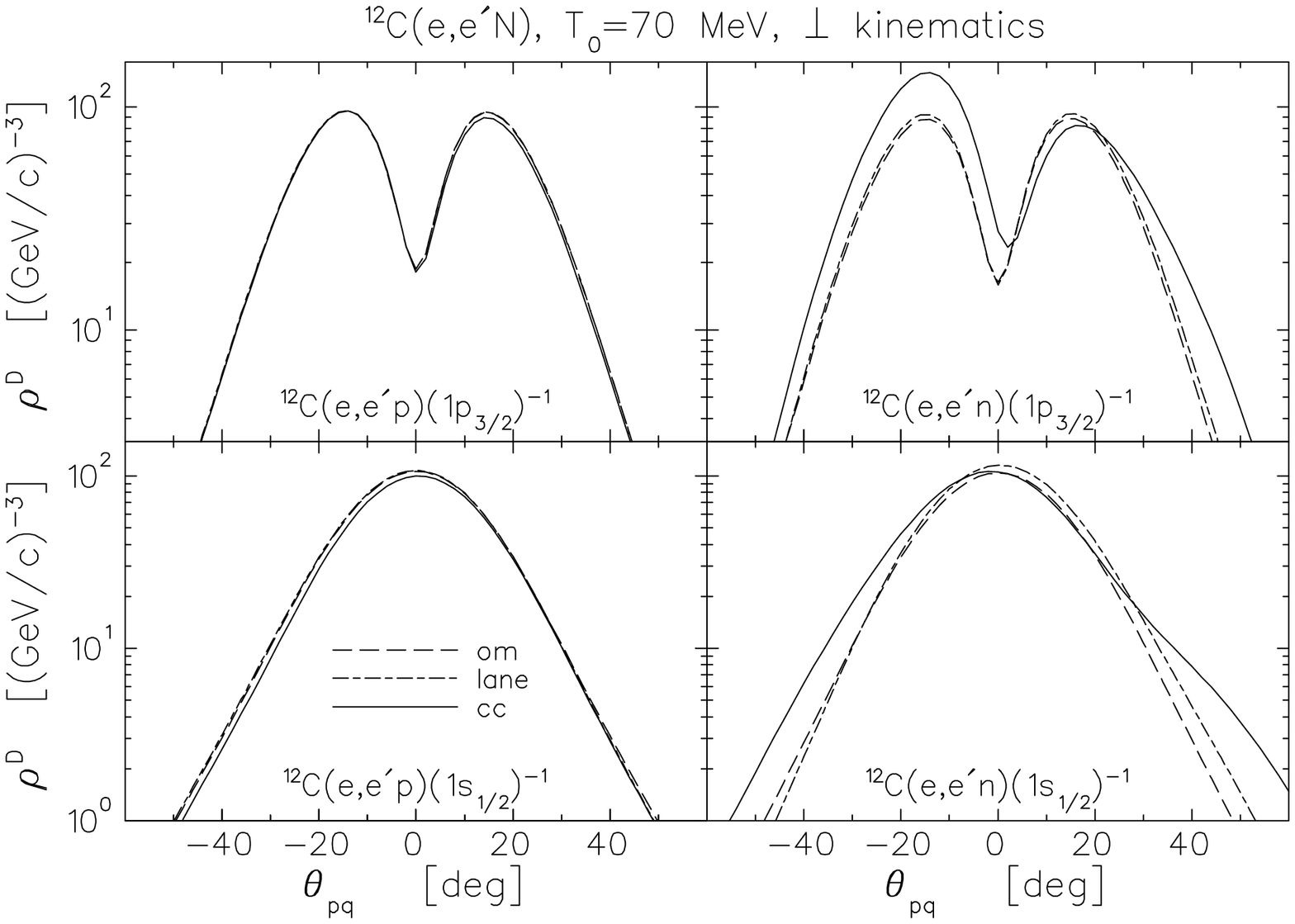,height=4.0in,angle=90} }
\caption{Distorted momentum distributions for $^{12}$C$(e,e^\prime N)$
in quasiperpendicular kinematics with $T_0 = 70$ MeV.
Dashed curves show the optical model (OM), dash-dotted curves the Lane
model, and solid curves the full coupled-channels calculation (CC).
These calculations are normalized to full subshell occupancy.}
\label{fig:perp_spectd_12C}
\end{figure}

\begin{figure}[htb] 
\centerline{
\strut\psfig{file=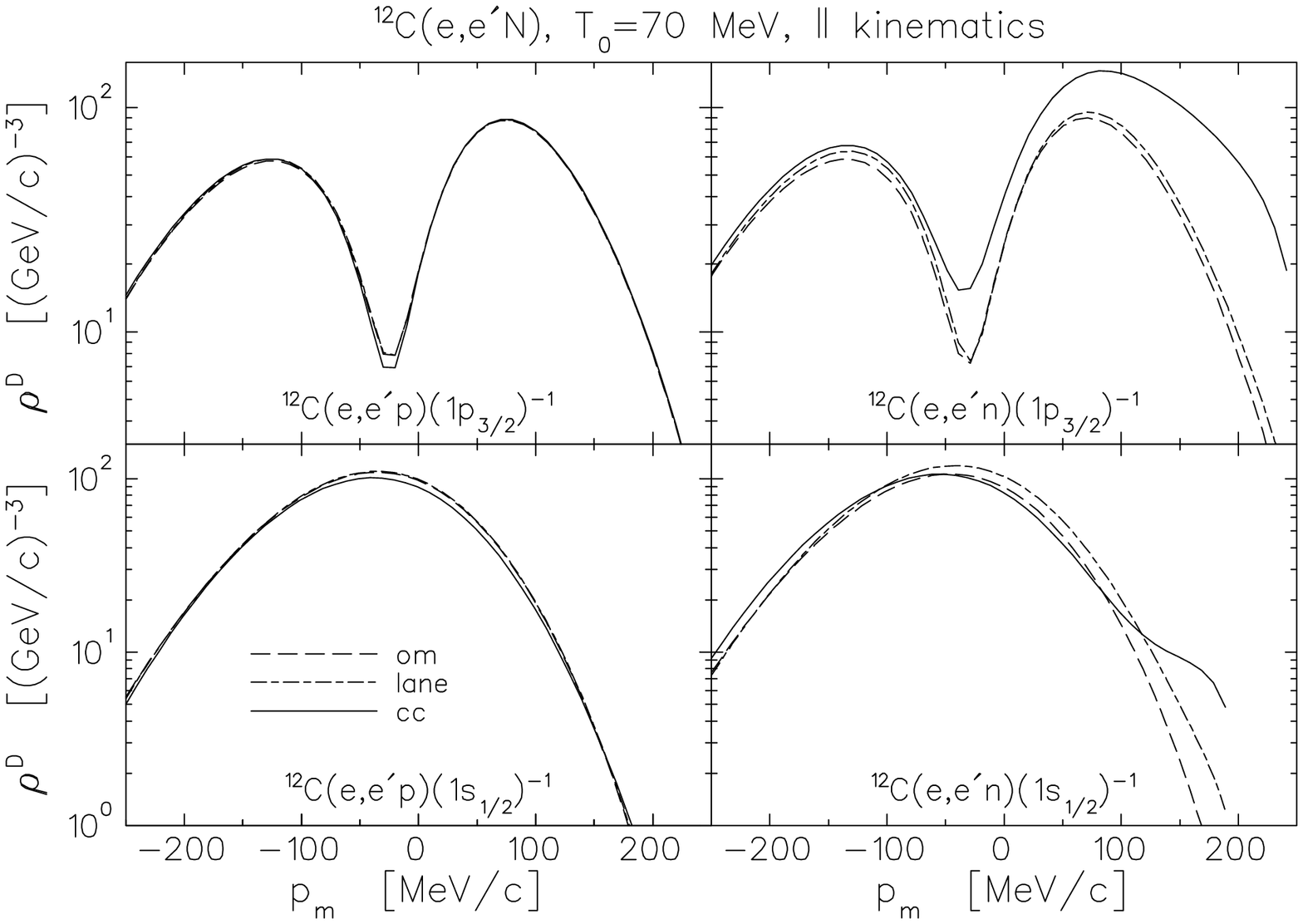,height=4.0in,angle=90} }
\caption{Distorted momentum distributions for $^{12}$C$(e,e^\prime N)$
in parallel kinematics with $T_0 = 70$ MeV.
See Fig.\ \protect{\ref{fig:perp_spectd_12C}} for legend.
}
\label{fig:para_spectd_12C}
\end{figure}

Distorted momentum distributions for $1p_{3/2}$ and $1s_{1/2}$ knockout 
are shown for
quasiperpendicular kinematics in Fig.\  \ref{fig:perp_spectd_12C} 
and for
parallel kinematics in Fig.\ \ref{fig:para_spectd_12C}.
These calculations are normalized to full subshell occupancy.
We find that charge exchange within the Lane model has rather little
effect, in qualitative agreement with Giusti and Pacati 
but in sharp disagreement with van der Steenhoven {\it et al}.
Furthermore, the more complete model of channel coupling suggests very
large contributions to $1p_{3/2}$ neutron knockout, in qualitative
agreement with Jeschonnek {\it et al}., 
who do employ a more complete representation of the nucleon-nucleon 
interaction in the final state.
The effects for proton knockout, especially for $1s_{1/2}$,
are not entirely negligible either.
These findings are independent of details of the kinematics, choice of 
optical potentials, or effective interactions, 
but are characteristic of the coupling schemes.
The Lane model only couples analog states via spin-independent central 
potentials, whereas the dominant isospin-changing final-state interaction 
at these energies is $t_{\sigma \tau}$, 
which includes both spin and isospin transfer and tends to
stimulate Gamow-Teller (GT) transitions.
We also find that coupling to the $1s_{1/2}$ hole states is very
important to $1p_{3/2}$ neutron knockout.

We also investigated the effect of expanding the model space to include
$1p_{1/2}$ configurations.
These states have relatively little effect upon the results shown here
whether or not direct knockout from  $1p_{1/2}$ orbitals is considered.

\begin{figure}[htb] 
\centerline{
\strut\psfig{file=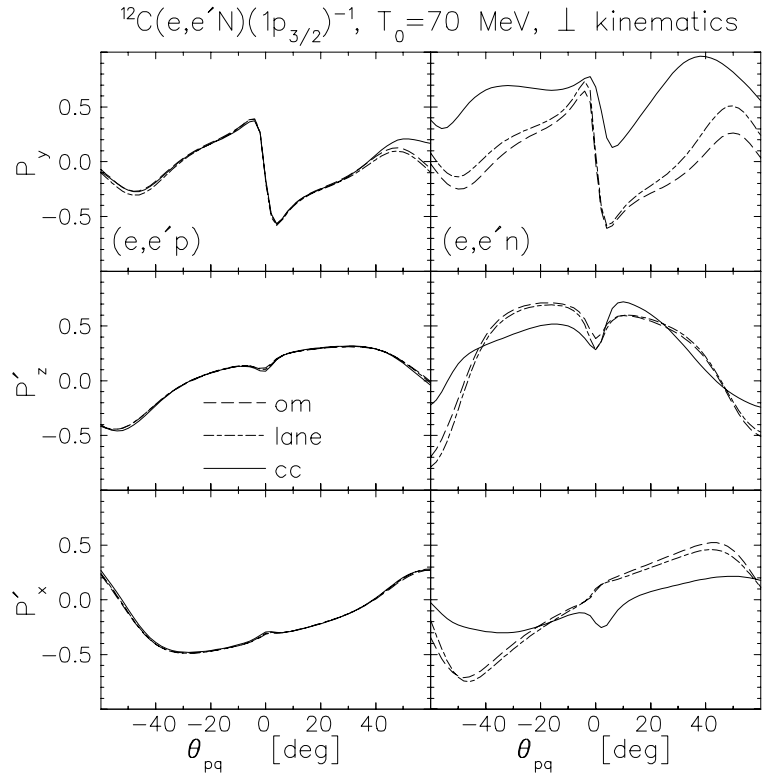,height=4.0in} }
\caption{Polarization of the recoil nucleon for $1p_{3/2}$ knockout in the
$^{12}$C$(\vec{e},e^\prime \vec{N})$ reaction
using quasiperpendicular kinematics with $T_0 = 70$ MeV.
Proton (neutron) knockout is shown in the left (right) column.
See Fig.\ \protect{\ref{fig:perp_spectd_12C}} for legend.
}
\label{fig:perp_pol_12C1p3}
\end{figure}

\begin{figure}[htb] 
\centerline{
\strut\psfig{file=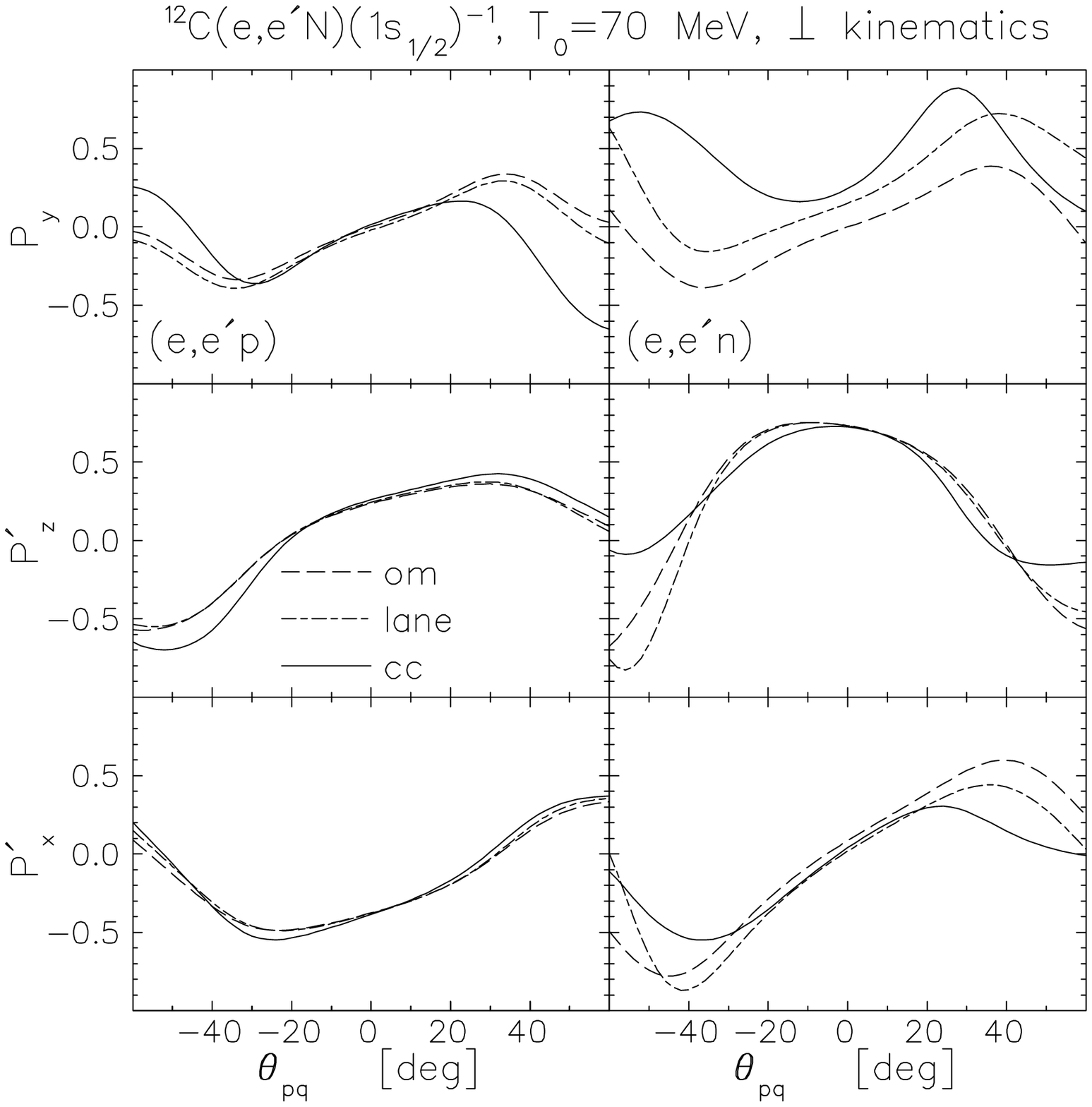,height=4.0in} }
\caption{Polarization of the recoil nucleon for $1s_{1/2}$ knockout in the
$^{12}$C$(\vec{e},e^\prime \vec{N})$ reaction
using quasiperpendicular kinematics with $T_0 = 70$ MeV.  
See Fig.\ \protect{\ref{fig:perp_pol_12C1p3}} for legend.
}
\label{fig:perp_pol_12C1s1}
\end{figure}

\begin{figure}[htb] 
\centerline{
\strut\psfig{file=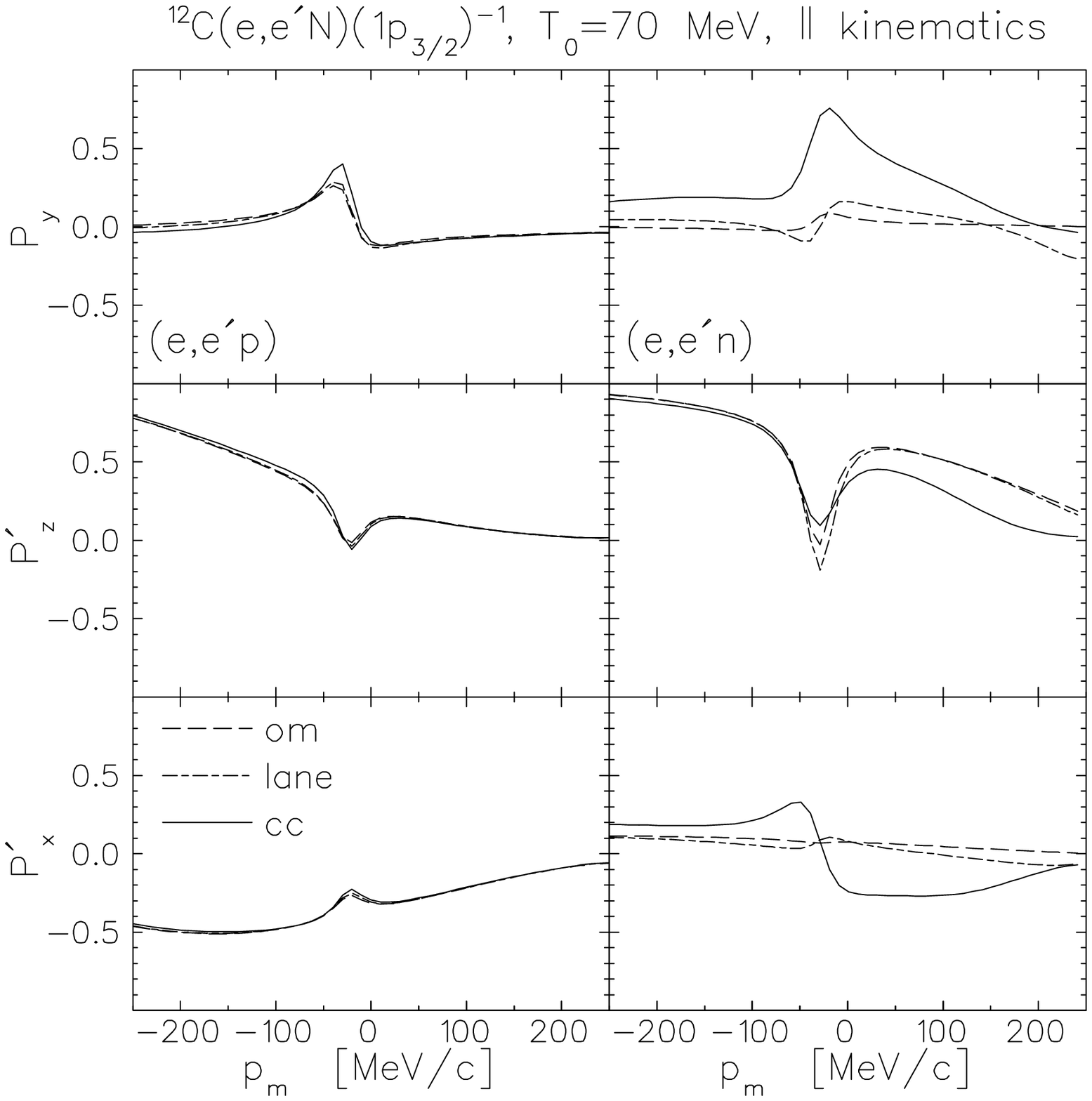,height=4.0in} }
\caption{Polarization of the recoil nucleon for $1p_{3/2}$ knockout in the
$^{12}$C$(\vec{e},e^\prime \vec{N})$ reaction
using parallel kinematics with $T_0 = 70$ MeV.
See Fig.\ \protect{\ref{fig:perp_pol_12C1p3}} for legend.
}
\label{fig:para_pol_12C1p3}
\end{figure}

\begin{figure}[htb] 
\centerline{
\strut\psfig{file=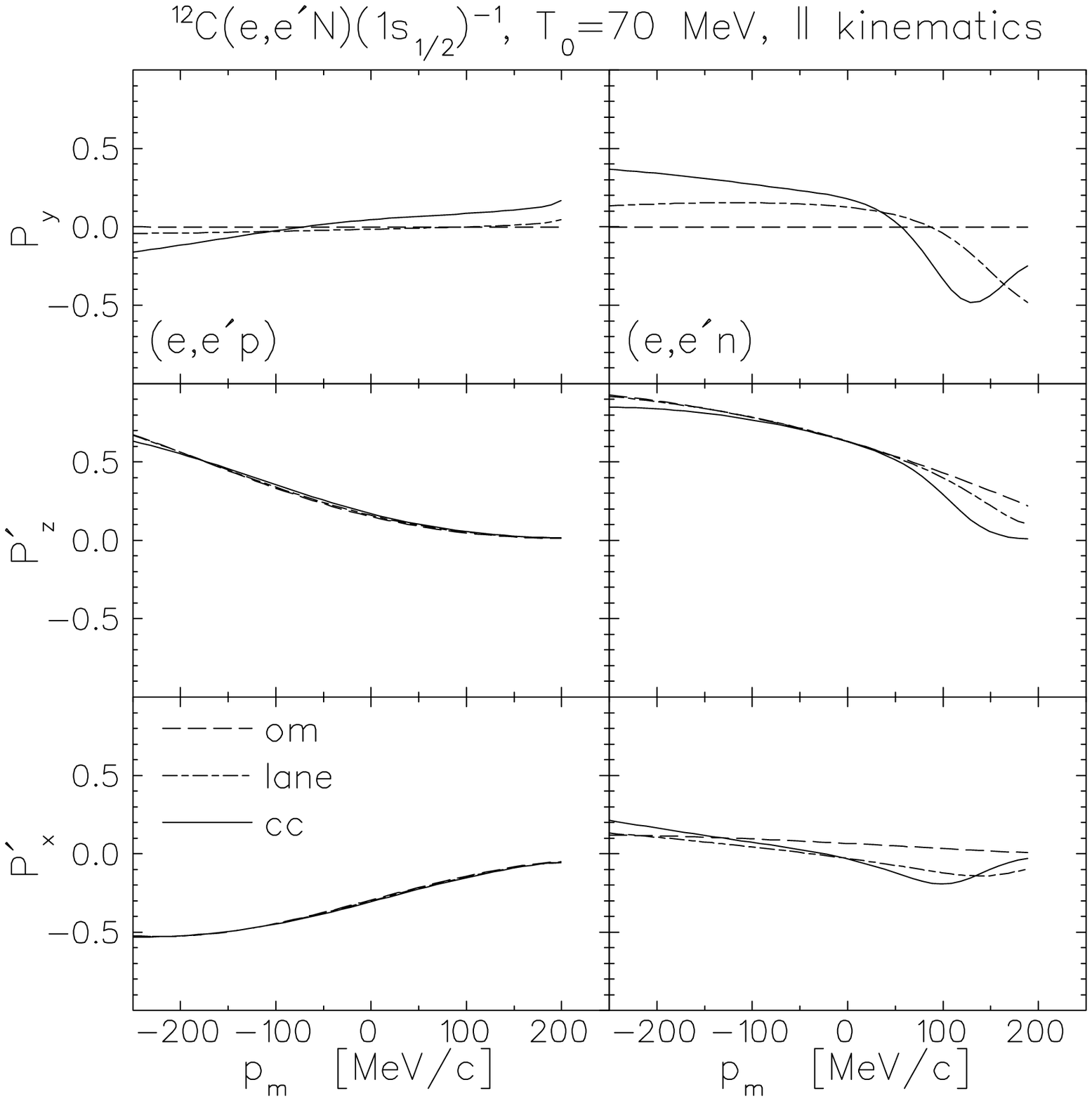,height=4.0in} }
\caption{Polarization of the recoil nucleon for $1s_{1/2}$ knockout in the
$^{12}$C$(\vec{e},e^\prime \vec{N})$ reaction
using parallel kinematics with $T_0 = 70$ MeV.
See Fig.\ \protect{\ref{fig:perp_pol_12C1p3}} for legend.
}
\label{fig:para_pol_12C1s1}
\end{figure}

In Figs.\ \ref{fig:perp_pol_12C1p3} -- \ref{fig:para_pol_12C1s1} 
we show recoil polarizations for nucleon knockout at
$T_0 = 70$ MeV expressed in the polarimeter basis.
The greatest sensitivity to channel coupling is seen in $P_y$, 
which is independent of electron helicity and vanishes without FSI.
The effects of channel coupling are much larger for neutron knockout than
for proton knockout, and much larger for the full model than for the Lane
model used by Giusti and Pacati \cite{Giusti89}.
Note that without channel coupling $P_y$ for analog states reached by
either neutron or proton knockout would be quite similar and that the
isospin differences produced by the Lane potential are fairly small, 
but that the spin-isospin final-state interaction $t_{\sigma\tau}$ produces
large differences between $P_y$ for neutron and proton knockout.

The helicity-dependent polarization components, $P^\prime_x$ and $P^\prime_z$,
do not require FSI and, hence, are less sensitive to channel coupling.
We have also shown that these quantities are relatively insensitive to
ambiguities in the single-nucleon current operator and
to the choice of optical model \cite{Kelly97}.
Thus, it has been proposed that the ratio $P^\prime_x / P^\prime_z$ is
sensitive to the form factor ratio $G_E / G_M$ in the nuclear medium.
Figs.\ \ref{fig:perp_pol_12C1p3} -- \ref{fig:para_pol_12C1s1}  
suggest that for proton knockout with modest missing momentum, 
channel coupling in FSI should not complicate this analysis either, 
even for these rather low ejectile energies.  
[Nevertheless, two-body currents beyond the scope of the present 
investigation may play an important role.]
However, for neutron knockout channel coupling does substantially affect 
the helicity-dependent recoil polarizations and, at least for this
energy regime, would complicate similar attempts to deduce neutron form 
factors in the nuclear medium. 
With a more complete model of the final-state interactions, we obtain 
a much larger coupled-channels effect on polarization transfer for
neutron knockout than calculated by Giusti and Pacati using the Lane model.

\subsection{Induced polarization}
\label{sec:12C-Pn}

\begin{figure}[htb] 
\centerline{
\strut\psfig{file=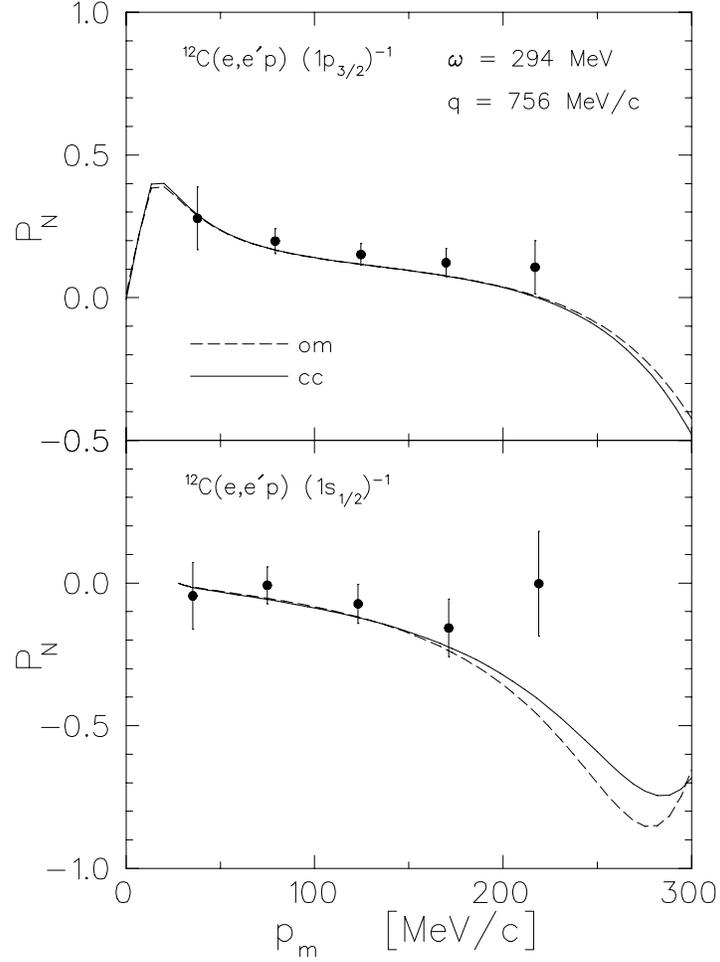,height=5.0in} }
\caption{Induced polarization of the recoil proton in the
$^{12}$C$(e,e^\prime \vec{p})$ reaction are compared with the data
of Woo \protect{\it et al.} \protect\cite{Woo98}.
The data for the s-shell are restricted to $28 < E_m < 39$ MeV to
limit the contribution of the underlying continuum. 
Dashed curves show the optical model (OM) and solid curves show the full 
coupled-channels calculation (CC).}
\label{fig:Woo}
\end{figure}

The first measurements of the induced polarization, $P_N$, for a nucleus 
with $A > 2$ were made by Woo {\it et al.}\ \cite{Woo98} 
for $^{12}$C$(e,e^\prime \vec{p})$
and the data for the $1p_{3/2}$ shell were found to be in good agreement 
with calculations based upon the distorted-wave impulse approximation
(DWIA) using phenomenological optical potentials fitted to proton 
scattering data.
However, it is important to test whether channel coupling affects the
induced polarization because $P_N$ would vanish without FSI. 
In Fig.\ \ref{fig:Woo} we compare calculations of the induced polarization for
$^{12}$C$(e,e^\prime \vec{p})$ with the recent data of 
Woo {\it et al}.\ \cite{Woo98} with $T_0 = 274$ MeV in 
quasiperpendicular kinematics.  
Final-state interactions were based upon the empirical effective interactions
tabulated in Ref. \cite{Kelly94a}, using linear interpolation with respect
to energy.
We find that channel coupling has very little effect upon the calculation for 
$1p_{3/2}$ knockout,
but is appreciable for $1s_{1/2}$ knockout when $p_m \gtrsim 200$ MeV/c;
unfortunately, the data do not extend far enough to test that effect.

\section{Channel coupling in $^{16}$O$(\lowercase{\vec{e},e}^\prime \vec{N})$ }
\label{sec:16O}

\subsection{Overlap functions}
\label{sec:overlap_16o}

\begin{figure}[htb] 
\centerline{
\strut\psfig{file=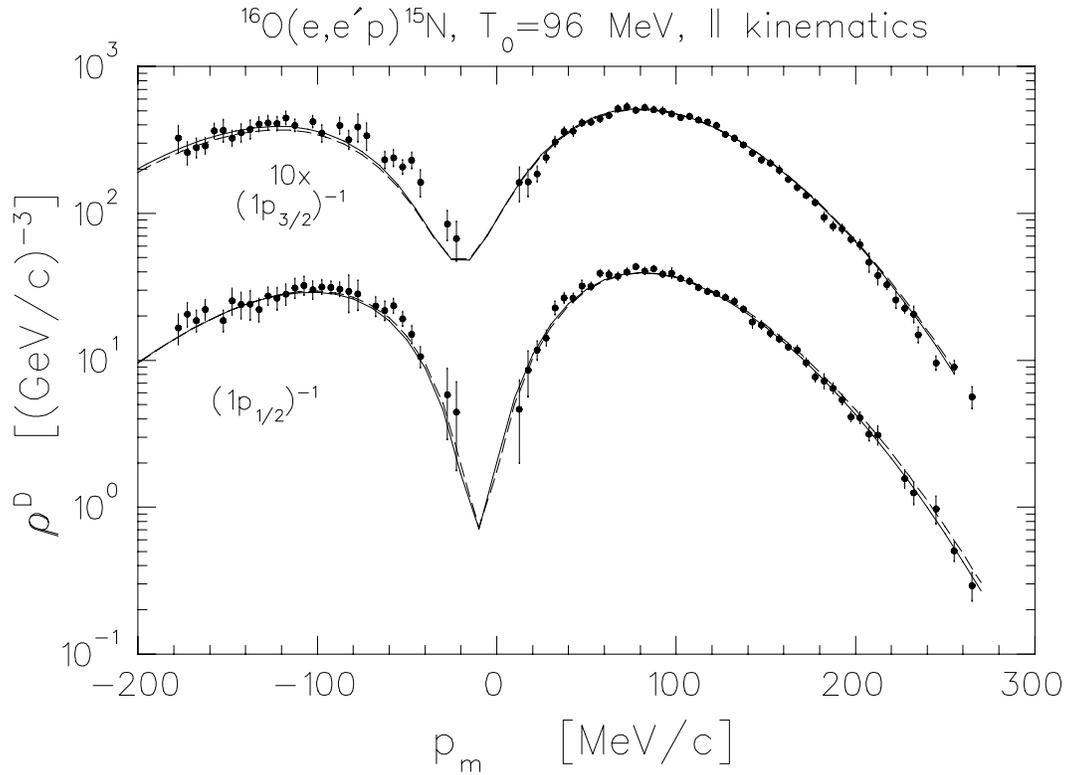,height=4.0in,angle=90} }
\caption{Distorted momentum distributions for $^{16}$O$(e,e^\prime p)$
in parallel kinematics with $T_0 = 96$ MeV.
Spectroscopic factors of 1.30 for $1p_{1/2}$ and 2.48 for $1p_{3/2}$ are
used to fit the calculations to the data of
Leuschner \protect{\it et al.} \protect\cite{Leuschner94} for the 
dominant p-shell fragments.
Dashed curves show the optical model (OM) and solid curves show the 
full coupled-channels calculation (CC).}
\label{fig:16o100}
\end{figure}

The overlap functions for p-shell proton knockout from $^{16}$O were 
obtained from the $^{16}$O$(e,e^\prime p)$ measurements of 
Leuschner {\it et al}.\ \cite{Leuschner94}.
The data for parallel kinematics with $T_0 = 96$ MeV are compared with
optical model calculations using the Paris-Melbourne effective interaction
in Fig.\ \ref{fig:16o100}.
Spectroscopic factors of 1.30 for $1p_{1/2}$ and 2.48
for $1p_{3/2}$ provide good visual fits to the data, 
but other choices of optical potential which also provide good descriptions
of proton elastic scattering can give spectroscopic factors which 
differ by 10\% or more \cite{Leuschner94,Kelly96}.
Coupled-channels calculations are shown also, but deviations of a few percent
are hardly visible on this scale.
For the s-shell we used the parametrization of Elton and Swift \cite{Elton67}.
For $^{15}$O we used the same potential shapes and adjusted the
central well depths to reproduce the separation energies for each state.

\subsection{Channel coupling in $^{16}$O$(\lowercase{e,e}^\prime N)$ at
$T_0 = 200$ MeV}
\label{sec:MAMI}

We begin by considering the kinematics of MAMI experiment
A1/2-93 \cite{A1/2-93},
which will measure the $^{16}$O$(\vec{e},e^\prime \vec{p})$ reaction
using quasiperpendicular kinematics with $E_0 = 855$ MeV, $\omega = 215$ MeV,
$q = 648$ MeV/c such that the ejectile energy is approximately 200 MeV.
Measurements of all three components of the recoil polarization will be made 
with polarized beam on both sides of ${\bf q}$ in order to separate
the even from the odd response functions.
Thus, if statistics permit, it should be possible to separate 
$R_{LT}$, $R^N_{LT}$, 
$R^{\prime L}_{LT}$,     $R^{\prime L}_{TT}$,
$R^{\prime S}_{LT}$, and $R^{\prime S}_{TT}$
for several opening angles $\theta_{pq}$.
For completeness, we have also performed calculations for parallel
kinematics using constant $T_0 = 200$ MeV.

Similar calculations for $T_0 = 135$ MeV were shown in Ref.\ \cite{Kelly96}.
Although the results are similar, the details depend upon ejectile energy.
Furthermore, the calculations of  Ref.\ \cite{Kelly96} did not include
$1s_{1/2}$ states in the model space, which we have since found to be
important.

\subsubsection{Distorted momentum distributions}
\label{sec:MAMI-spectd}

\begin{figure}[htb] 
\centerline{
\strut\psfig{file=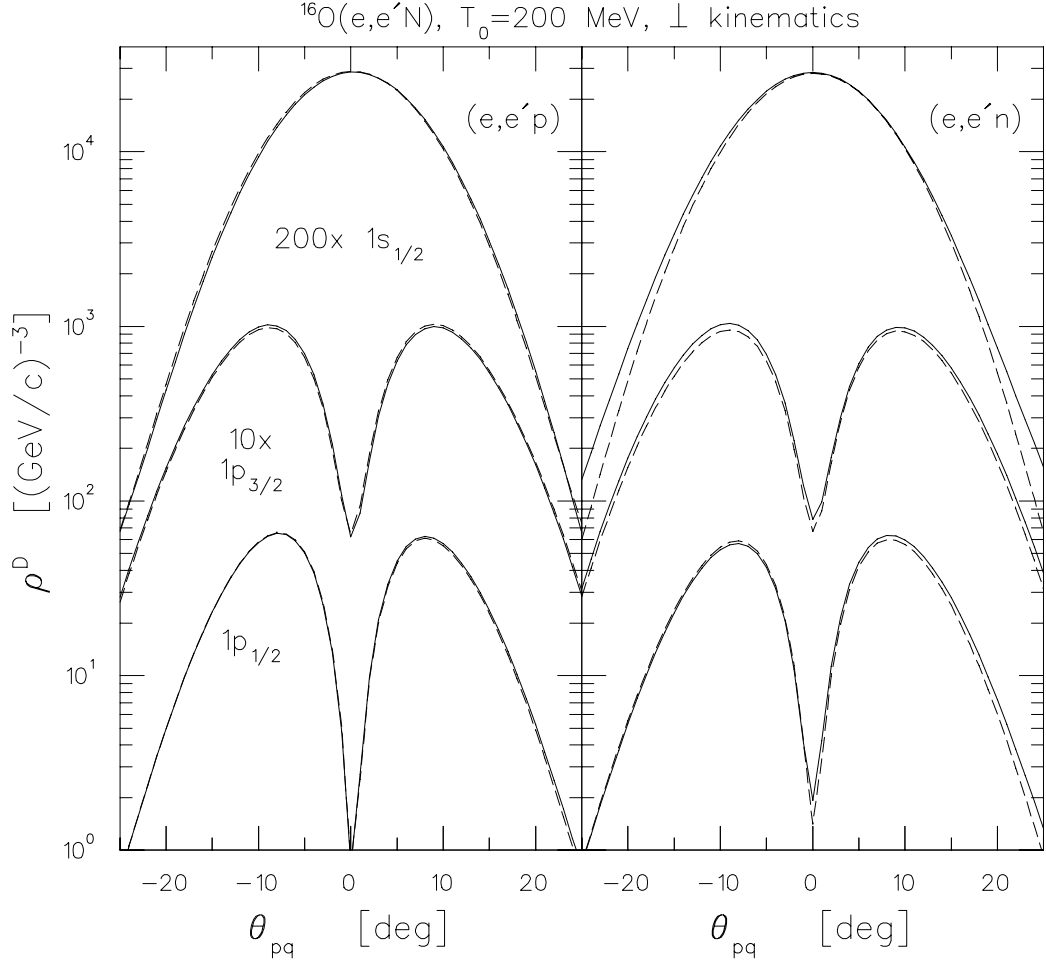,height=5.0in,angle=90} }
\caption{Distorted momentum distributions for $^{16}$O$(e,e^\prime N)$
in quasiperpendicular kinematics with $T_0 = 200$ MeV.
Proton (neutron) knockout is shown on the left (right) side.
These calculations are normalized to full subshell occupancy.
See Fig.\ \protect{\ref{fig:16o100}} for legend.
}
\label{fig:perp_spectd_16O_Mainz}
\end{figure}

\begin{figure}[htb] 
\centerline{
\strut\psfig{file=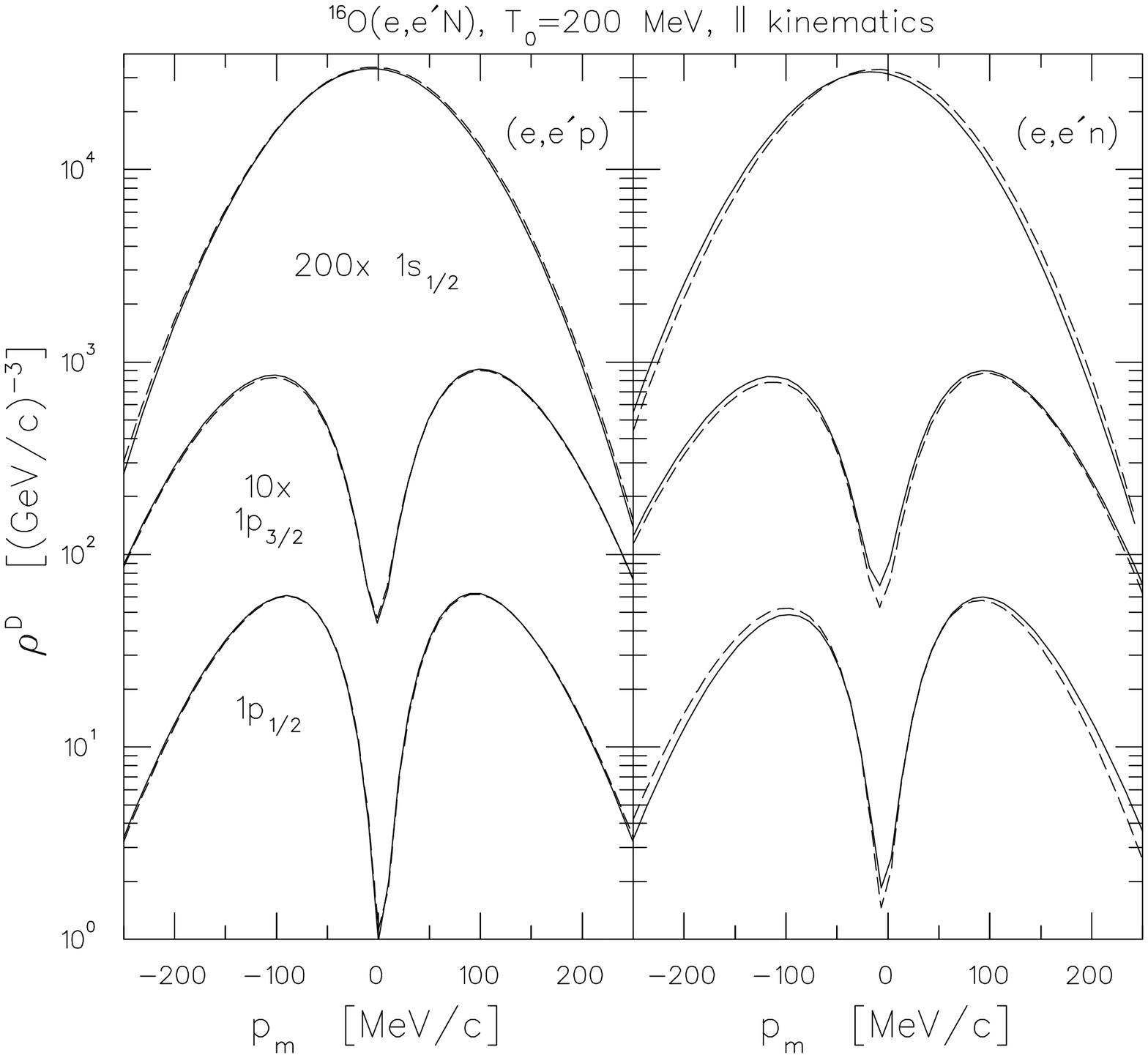,height=5.0in,angle=90} }
\caption{Distorted momentum distributions for $^{16}$O$(e,e^\prime N)$
in parallel kinematics with $T_0 = 200$ MeV.
See Fig.\ \protect{\ref{fig:perp_spectd_16O_Mainz}} for legend.
}
\label{fig:para_spectd_16O_Mainz}
\end{figure}

\begin{figure}[htb] 
\centerline{
\strut\psfig{file=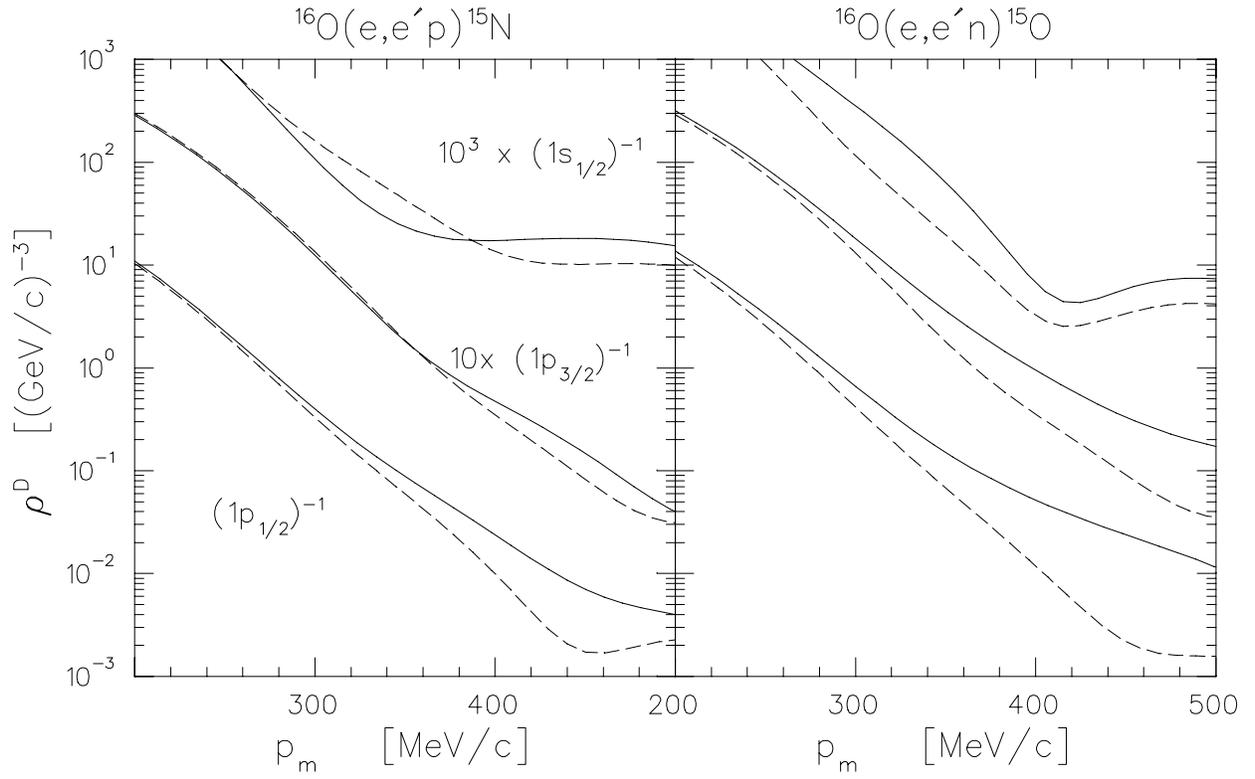,height=4.0in,angle=90} }
\caption{Distorted momentum distributions for $^{16}$O$(e,e^\prime N)$
in quasiperpendicular kinematics with $T_0 = 200$ MeV, selecting
large missing momenta for $\theta_p > \theta_q$.
See Fig.\ \protect{\ref{fig:perp_spectd_16O_Mainz}} for legend.
}
\label{fig:large_pm_16O_Mainz}
\end{figure}

The reduced cross sections are shown in 
Fig.\ \ref{fig:perp_spectd_16O_Mainz} for quasiperpendicular kinematics 
and in
Fig.\ \ref{fig:para_spectd_16O_Mainz} for parallel kinematics.
In addition, large missing momenta for quasiperpendicular kinematics with
$\theta_p > \theta_q$ are shown in
Fig.\ \ref{fig:large_pm_16O_Mainz}.
These calculations are normalized to full subshell occupancy.
The effect of channel coupling upon the reduced cross section for
proton knockout appears to be quite small for $p_m \lesssim 300$ MeV/c, 
but can become appreciable for large $p_m$.
At this energy channel coupling enhances the calculated cross section for 
p-shell proton knockout by factors of approximately 1.5 -- 2 for 
$p_m \sim 500$ MeV/c
and significantly alters the shape of the missing momentum distribution 
for $1s_{1/2}$ proton knockout.
Similar calculations for $T_0 = 135$ MeV \cite{Kelly96} show larger
factors, especially for the $1p_{1/2}$ state, but details of these
effects depend upon ejectile energy. 
Substantially larger enhancements of the cross section for $p_m > 300$
MeV/c were predicted for the rotational band in $^{10}$B$(e,e^\prime p)^9$Be, 
with both reorientation and inelastic scattering being equally important
\cite{deBever94}, because the quadrupole coupling is larger for that
strongly deformed system.
Hence, we conclude that the relative importance of various final-state 
interaction mechanisms depends upon nuclear structure in an essential manner.
Furthermore, such effects will need to be examined carefully before any
conclusions about high momentum components due to short-range correlations
are drawn from proton knockout data.

For $^{16}$O$(e,e^\prime n)^{15}$O, channel coupling is significant even
for $p_m$ near the peaks of the momentum distributions.
The most important couplings are those which change both the spin and the 
isospin of the residual nucleus.
Although the effect on the cross section for quasiperpendicular kinematics
is relatively small, changes in the left-right asymmetries for p-shell
neutron knockout reflect substantial changes in the $R_{LT}$ response 
functions that arise primarily from charge-exchange in FSI.
Similarly, the modest enhancements of the cross section for neutron
knockout in parallel kinematics originate in charge-exchange contributions
to the longitudinal form factor.
We also find that the missing momentum distributions for s-shell neutron
knockout are broadened in quasiperpendicular and shifted in parallel
kinematics by charge exchange.
For $p_m > 300$ MeV/c channel coupling enhances the cross section for
neutron knockout by large factors relative to the conventional
optical-model calculation.
For $p_m \sim 500$ MeV/c and $T_0 = 200$ MeV, 
these factors approach an order of magnitude for p-shell neutron knockout 
with both charge exchange and inelastic scattering playing important roles.
The effect of charge exchange upon neutron knockout was much larger at
70 MeV than it is at 200 MeV.
Thus, we conclude that the importance of channel coupling decreases 
fairly rapidly as the ejectile energy increases, 
but for neutron knockout remains significant at 200 MeV.

\subsubsection{Recoil polarization}
\label{sec:MAMI-polarization}

\begin{figure}[htb] 
\centerline{
\strut\psfig{file=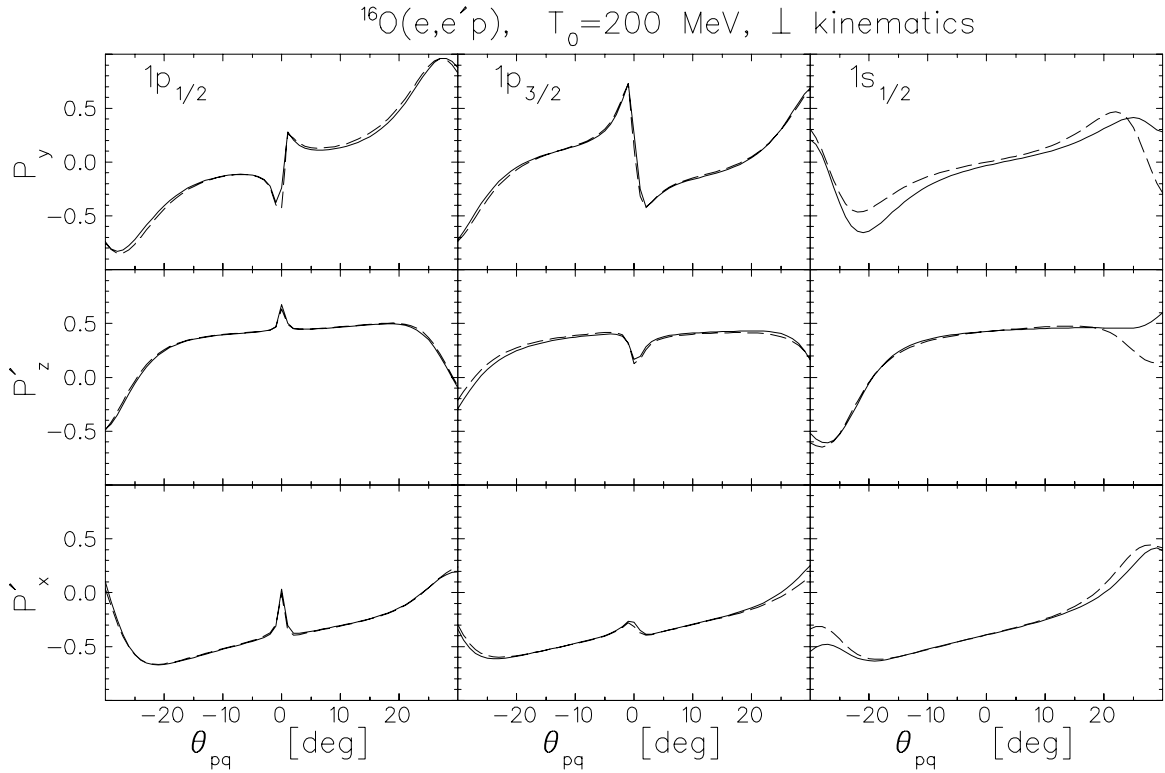,height=4.0in,angle=90} }
\caption{Polarization of the recoil proton in the
$^{16}$O$(\vec{e},e^\prime \vec{p})^{15}$N reaction
using quasiperpendicular kinematics with $T_0 = 200$ MeV.
The three columns show calculations for $1p_{1/2}$, $1p_{3/2}$, and
$1s_{1/2}$ knockout.
Dashed curves show the optical model (OM) and solid curves show the full 
coupled-channels calculation (CC).}
\label{fig:perp_pol_15N_Mainz}
\end{figure}

\begin{figure}[htb] 
\centerline{
\strut\psfig{file=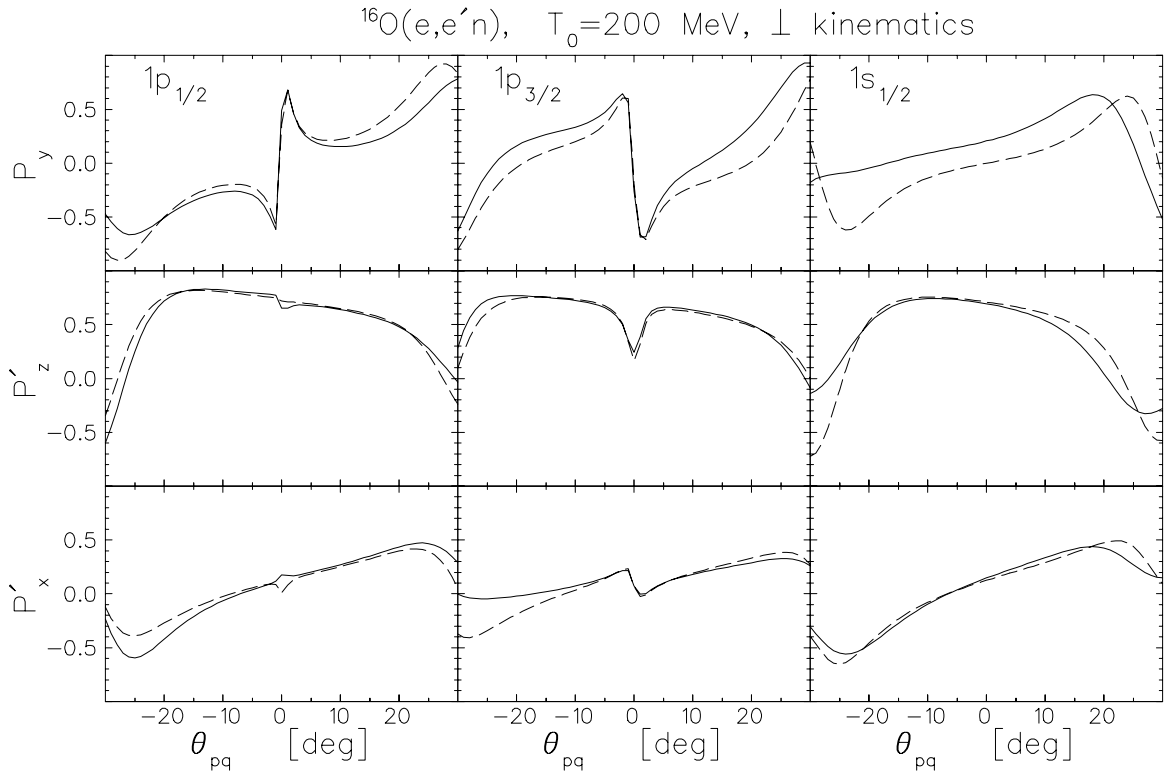,height=4.0in,angle=90} }
\caption{Polarization of the recoil neutron in the
$^{16}$O$(\vec{e},e^\prime \vec{n})^{15}$O reaction
using quasiperpendicular kinematics with $T_0 = 200$ MeV.
See Fig.\ \protect{\ref{fig:perp_pol_15N_Mainz}} for legend.
}
\label{fig:perp_pol_15O_Mainz}
\end{figure}

\begin{figure}[htb] 
\centerline{
\strut\psfig{file=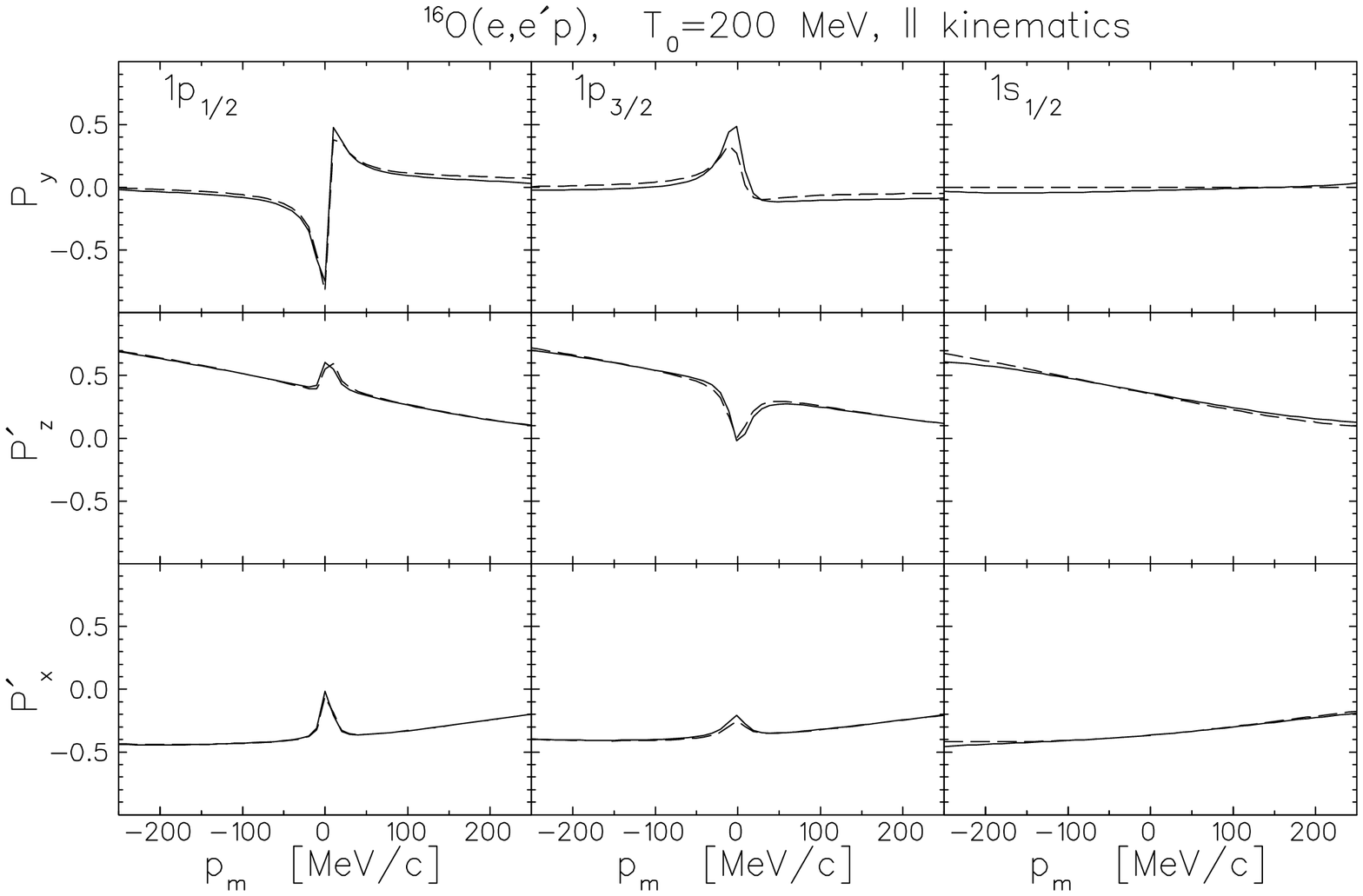,height=4.0in,angle=90} }
\caption{Polarization of the recoil proton in the
$^{16}$O$(\vec{e},e^\prime \vec{p})^{15}$N reaction
using parallel kinematics with $T_0 = 200$ MeV.
See Fig.\ \protect{\ref{fig:perp_pol_15N_Mainz}} for legend.
}
\label{fig:para_pol_15N_Mainz}
\end{figure}

\begin{figure}[htb] 
\centerline{
\strut\psfig{file=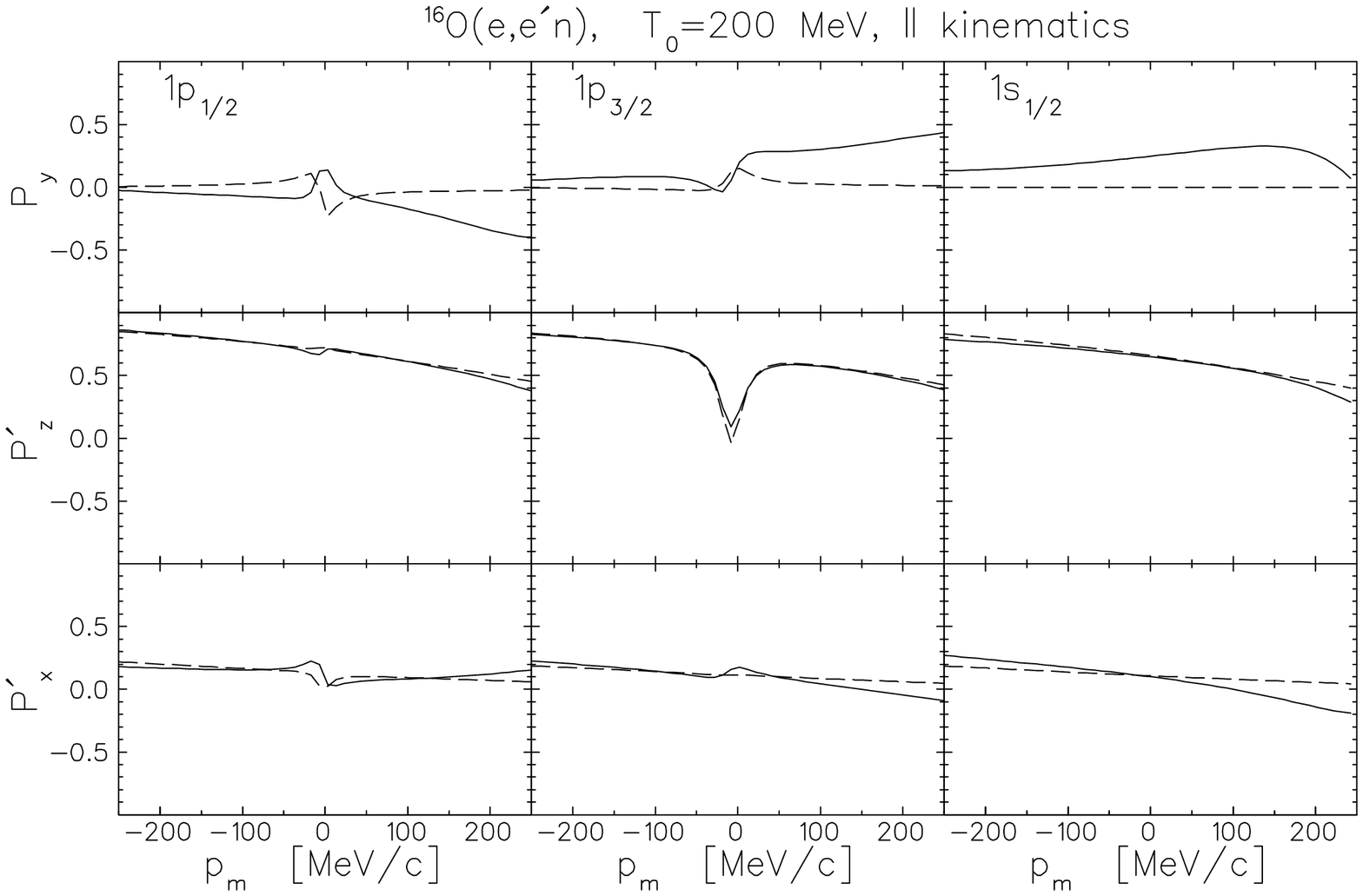,height=4.0in,angle=90} }
\caption{Polarization of the recoil neutron in the
$^{16}$O$(\vec{e},e^\prime \vec{n})^{15}$O reaction
using parallel kinematics with $T_0 = 200$ MeV.
See Fig.\ \protect{\ref{fig:perp_pol_15N_Mainz}} for legend.
}
\label{fig:para_pol_15O_Mainz}
\end{figure}

Recoil-polarization observables expressed in the polarimeter basis 
are shown in 
Fig.\ \ref{fig:perp_pol_15N_Mainz} -- \ref{fig:perp_pol_15O_Mainz}
for quasiperpendicular kinematics and in
Fig.\ \ref{fig:para_pol_15N_Mainz} -- \ref{fig:para_pol_15O_Mainz} 
for parallel kinematics.
For $p_m \lesssim 300$ MeV/c we find that channel coupling has 
practically no effect upon the polarization transfer for proton knockout.
The effects of channel coupling upon the polarization transfer for 
neutron knockout are much smaller than at $T_0 = 70$ MeV, 
but remain nonnegligible.  
Larger effects are obtained for $p_m \gtrsim 300$ MeV/c, 
but these variations remain comparable to those arising from ambiguities
in the off-shell current operator explored in Ref.\ \cite{Kelly97}.

The induced polarization, $P_y$, is found to be more sensitive to
channel coupling within final-state interactions.
Small but nonnegligible sensitivity to channel coupling 
in quasiperpendicular kinematics is exhibited by proton knockout,
particularly for the s-shell,
whereas for neutron knockout channel coupling remains quite important
even for modest missing momenta.
The induced polarization for parallel kinematics is quite small for
proton knockout, but for neutron knockout is substantially enhanced 
by channel coupling.
We also find that channel coupling is generally more important than
variations due to the choice of optical potential.
Furthermore, although these effects decrease as $T_0$ increases from
135 to 200 MeV, the energy dependence is fairly slow.  

It is interesting to note that $P_y$ for s-shell knockout in parallel 
kinematics vanishes without channel coupling in the final state, but that a
small polarization results from the spin dependence of channel coupling.
The presence of an underlying continuum would make it difficult 
to observe this effect for $1s_{1/2}$ knockout, 
but it should be possible to observe this polarization for isolated
s-shell knockout, such as $2s_{1/2}$ knockout from $^{40}$Ca, given
sufficient resolution.

\subsubsection{Response functions}
\label{sec:response-functions}

\begin{figure}[htb] 
\centerline{
\strut\psfig{file=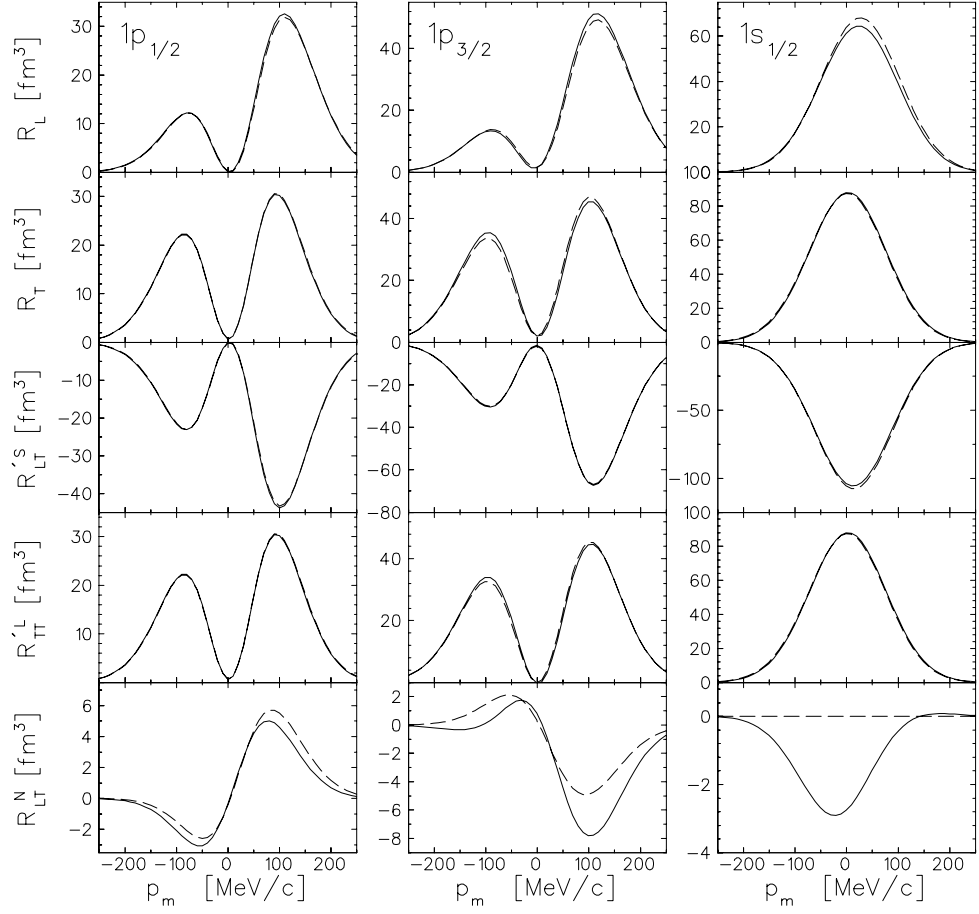,height=5.0in} }
\caption{Response functions for the
$^{16}$O$(\vec{e},e^\prime \vec{p})^{15}$N reaction
using parallel kinematics with $T_0 = 200$ MeV.
These calculations are normalized to full subshell occupancy.
See Fig.\ \protect{\ref{fig:perp_pol_15N_Mainz}} for legend.
}
\label{fig:para_rsfns_15N_Mainz}
\end{figure}

\begin{figure}[htb] 
\centerline{
\strut\psfig{file=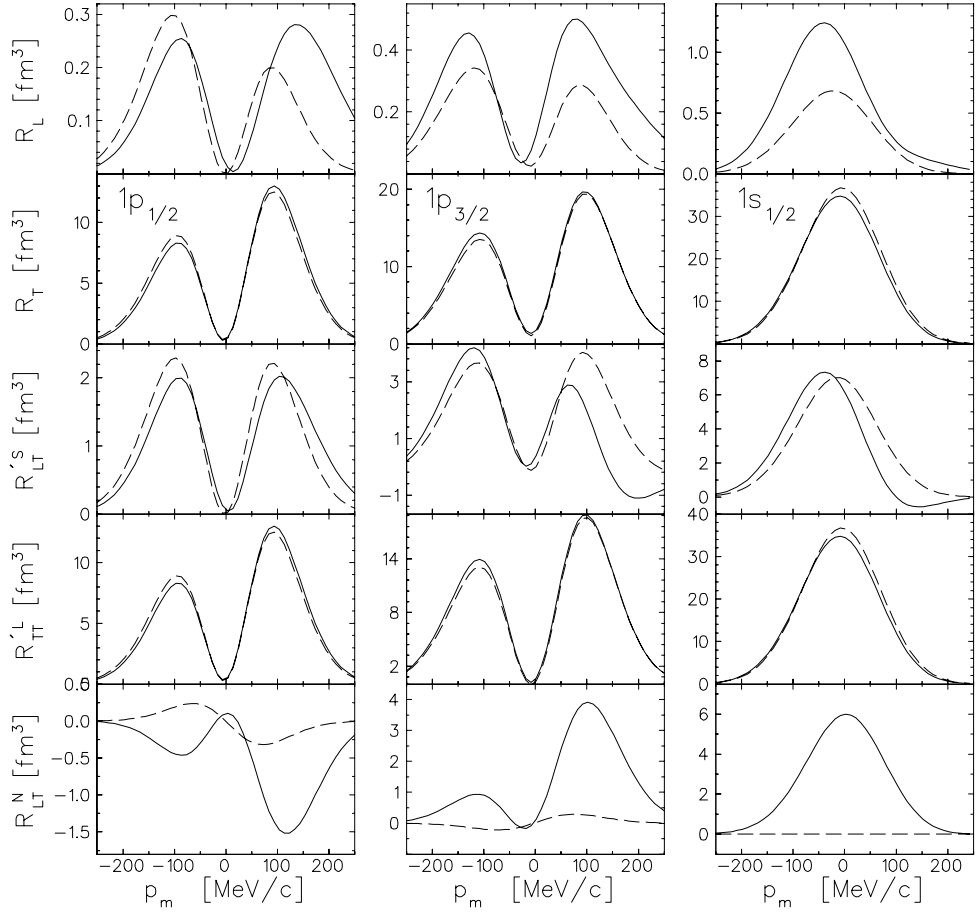,height=5.0in} }
\caption{Response functions for the
$^{16}$O$(\vec{e},e^\prime \vec{n})^{15}$O reaction
using parallel kinematics with $T_0 = 200$ MeV.
See Fig.\ \protect{\ref{fig:para_rsfns_15N_Mainz}} for legend.
}
\label{fig:para_rsfns_15O_Mainz}
\end{figure}

Response functions for parallel kinematics are shown in 
Fig.\ \ref{fig:para_rsfns_15N_Mainz} for 
$^{16}$O$(\vec{e},e^\prime \vec{p})^{15}$N
and in 
Fig. \ref{fig:para_rsfns_15O_Mainz} for 
$^{16}$O$(\vec{e},e^\prime \vec{n})^{15}$O.
These calculations are normalized to full subshell occupancy.
For proton knockout the largest effects are seen in $R_{LT}^N$, which
vanishes without FSI and for the s-shell vanishes without channel coupling;
hence, $R_{LT}^N$, which corresponds to $P_N$ for parallel kinematics,
tends to be most sensitive to details of the final-state interactions.
For neutron knockout we find that charge exchange in the final state
strongly enhances both $R_L$ and $R_{LT}^N$ and also has significant effects 
upon $R^{\prime S}_{LT}$,
whereas the effects upon purely transverse response functions are much 
smaller.

\begin{figure}[htb] 
\centerline{
\strut\psfig{file=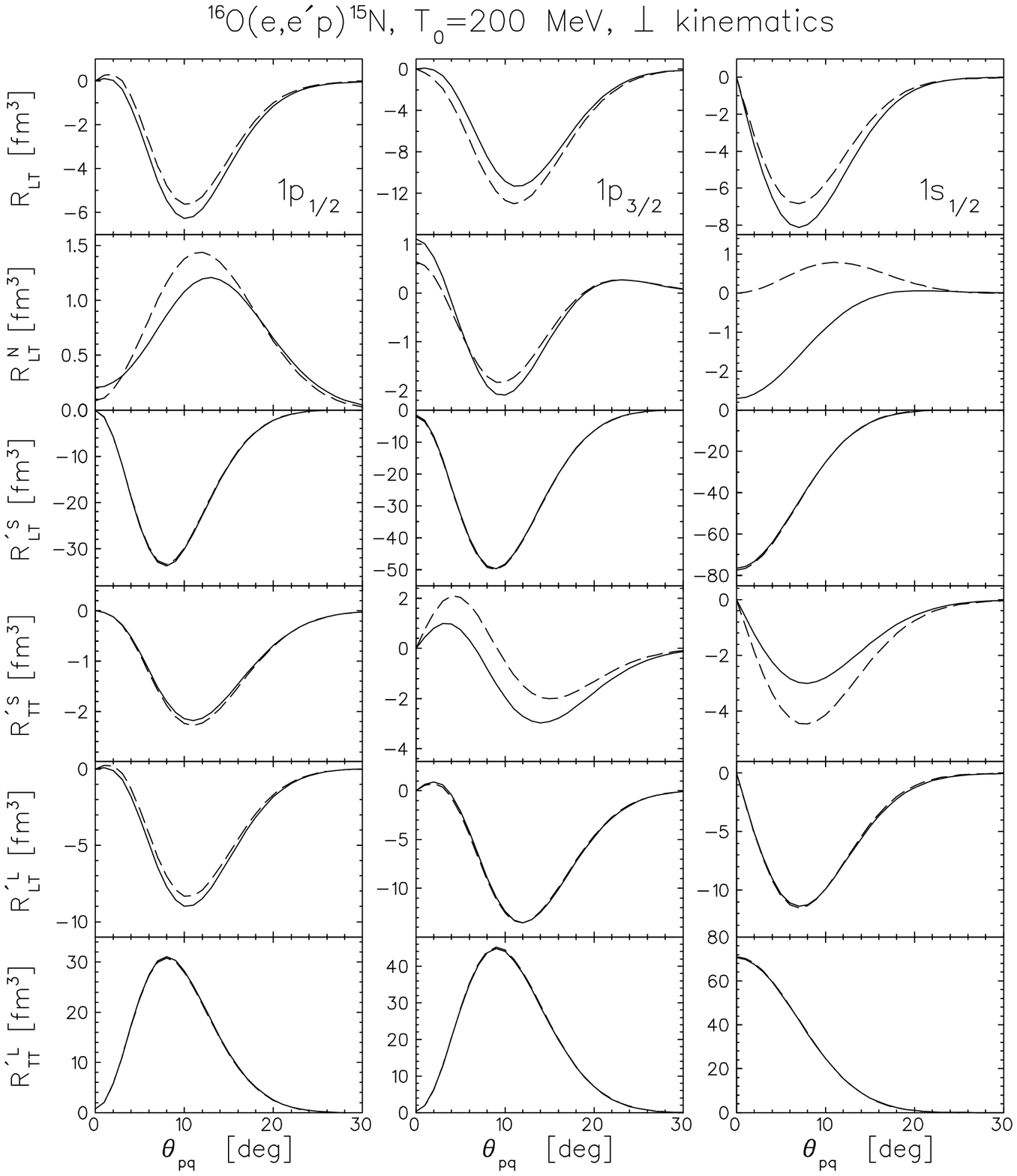,height=5.0in} }
\caption{Selected response functions for the
$^{16}$O$(\vec{e},e^\prime \vec{p})^{15}$N reaction
using coplanar quasiperpendicular kinematics with $T_0 = 200$ MeV.
See Fig.\ \protect{\ref{fig:para_rsfns_15N_Mainz}} for legend.
}
\label{fig:perp_rsfns_15N_Mainz}
\end{figure}

\begin{figure}[htb] 
\centerline{
\strut\psfig{file=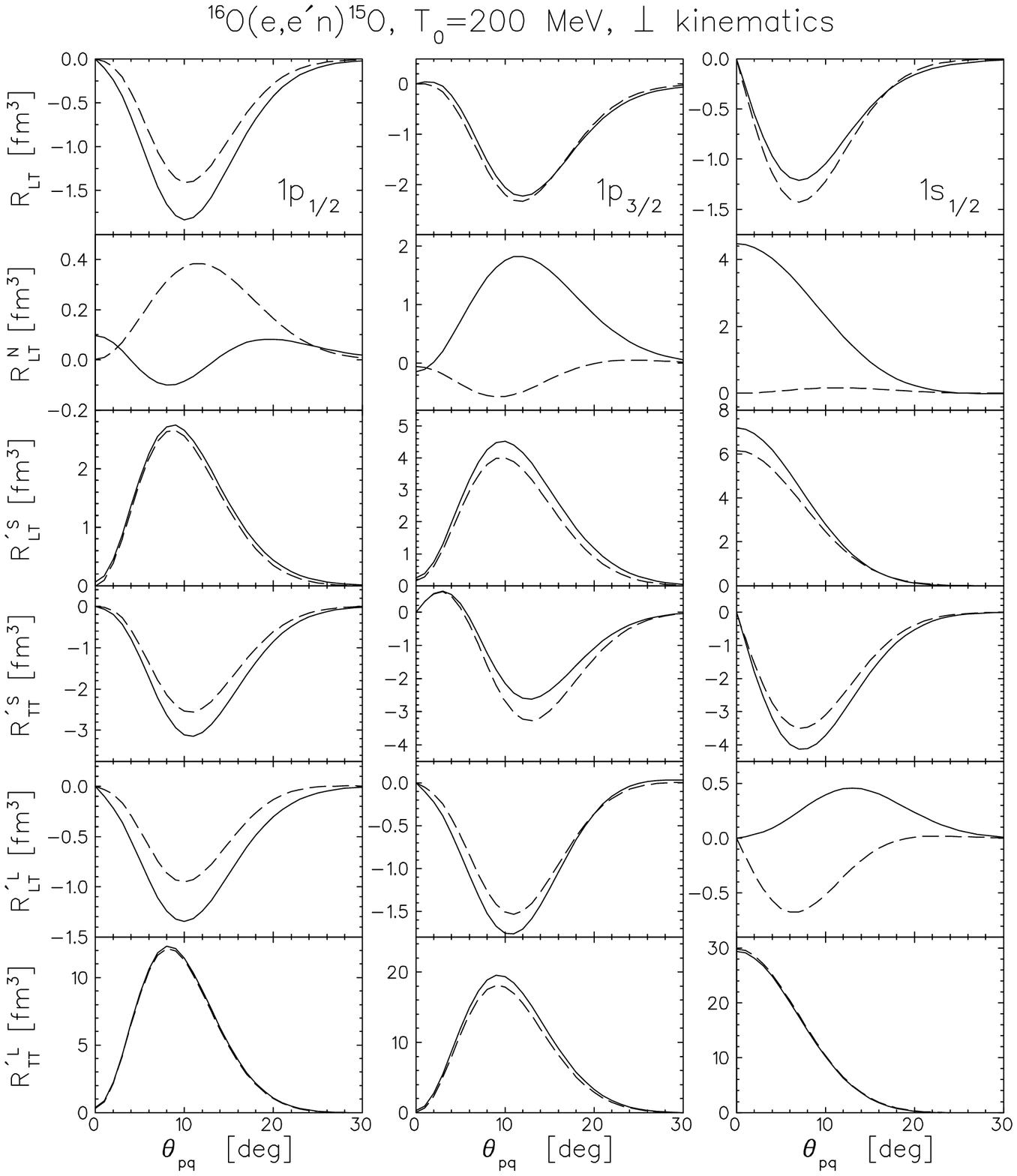,height=5.0in} }
\caption{Selected response functions for the
$^{16}$O$(\vec{e},e^\prime \vec{n})^{15}$O reaction
using coplanar quasiperpendicular kinematics with $T_0 = 200$ MeV.
See Fig.\ \protect{\ref{fig:para_rsfns_15N_Mainz}} for legend.
}
\label{fig:perp_rsfns_15O_Mainz}
\end{figure}

Selected response functions for coplanar quasiperpendicular kinematics 
are shown in 
Fig.\ \ref{fig:perp_rsfns_15N_Mainz} for 
$^{16}$O$(\vec{e},e^\prime \vec{p})^{15}$N
and in 
Fig. \ref{fig:perp_rsfns_15O_Mainz} for 
$^{16}$O$(\vec{e},e^\prime \vec{n})^{15}$O.
We chose to show those response functions which potentially can be deduced 
from cross section and recoil polarization measurements on both sides of
the momentum transfer vector for fixed electron scattering kinematics
because it is anticipated that MAMI experiment A1/2-93 \cite{A1/2-93} 
will provide data of this type.
The effects of channel coupling upon most strong Class I response functions
for proton knockout are relatively small, but at this energy remain
appreciable for $R_{LT}$ and $R_{TT}$.
For p-shell proton knockout opposite effects upon $R_{LT}$ are predicted
for the two spin-orbit partners.
There can also be significant effects upon some of the polarized Class I 
response functions for proton knockout, 
such as $R^{\prime S}_{TT}$ for the $1p_{3/2}$ and $1s_{1/2}$ states.
Therefore, the interpretation of response functions expected from
MAMI experiment A1/2-93 \cite{A1/2-93} will need to consider channel 
coupling in the final state.
The effects upon many of the Class II response functions, such as $R^N_{LT}$,
can be quite large even for proton knockout, especially for the
$1s_{1/2}$ state.
Note that without channel coupling, $R^m_{LT}$ and  $R^m_{TT}$ with 
$m \in \{L,S\}$ would vanish for s-shell knockout, 
but those response functions become appreciable when spin-dependent 
channel coupling is present in the final state.
Although not shown, strong modifications of $R^S_{LT}$,  $R^L_{LT}$,
and $R^L_{TT}$ are predicted for p-shell proton knockout also.

Most Class II response functions for neutron knockout are very strongly
affected by channel coupling, with the most important channel
couplings involving both isospin and angular momentum transfer.
Although not shown in these figures, the Lane model produces much smaller
effects because it lacks important spin-dependent and noncentral interactions.
Unlike proton knockout, many of the Class I response functions for
neutron knockout also exhibit substantial sensitivity to channel coupling, 
especially the  L- and LT-type response functions.
Stronger effects were obtained in calculations for $T_0 = 135$ MeV
where it was proposed to investigate the role of isobar currents in
$R_{LT}$ for $(e,e^\prime n)$. 
Although the relative importance of channel coupling decreases as the ejectile
energy increases, these effects remain important at $T_0 = 200$ MeV.
Therefore, it appears that it will be difficult to separate the effects
of two-body currents from those of FSI using neutron knockout at these
energies.
Perhaps higher ejectile energies will prove to be more favorable, 
but calculations including two-body currents are not available for larger
$Q^2$.

\subsection{Channel coupling in 
$^{16}$O$(\lowercase{\vec{e},e}^\prime \vec{N})$ at $T_0 = 433$ MeV}
\label{sec:TJNAF}

The first experiment to measure recoil polarization for polarized
electron scattering from a target with $A > 2$ was performed recently
at Jefferson Laboratory \cite{TJNAF89-033} and the data are 
presently being analyzed.
The experiment used $^{16}$O$(\vec{e},e^\prime \vec{p})$ in
quasiperpendicular kinematics with 
$E_0 = 2.445$ GeV, $\omega = 0.445$ GeV, and $q = 1.0$ GeV/c. 
Calculations for this reaction at $T_0 = 433$ MeV show that the effects of 
channel coupling on recoil polarization continue to decline as the 
ejectile energy increases;
these effects are similar to but smaller than those shown for 200 MeV.
Thus, it should be feasible to investigate possible medium modifications 
of the nucleon electromagnetic form factors for large $Q^2$ using 
quasifree recoil polarization.

A more quantitative assessment of the sensitivity of recoil polarization
to various aspects of the model, including final-state interactions, can
be made in terms of the polarization ratio
\begin{equation}
r_{xz} = P^\prime_x / P^\prime_z
\end{equation}
which for a free nucleon at rest is proportional to $G_E/G_M$.
We can then compare $r_{xz}$ for a particular model either to a 
plane-wave calculation or to a baseline optical-model calculation.

\begin{figure}[htb] 
\centerline{
\strut\psfig{file=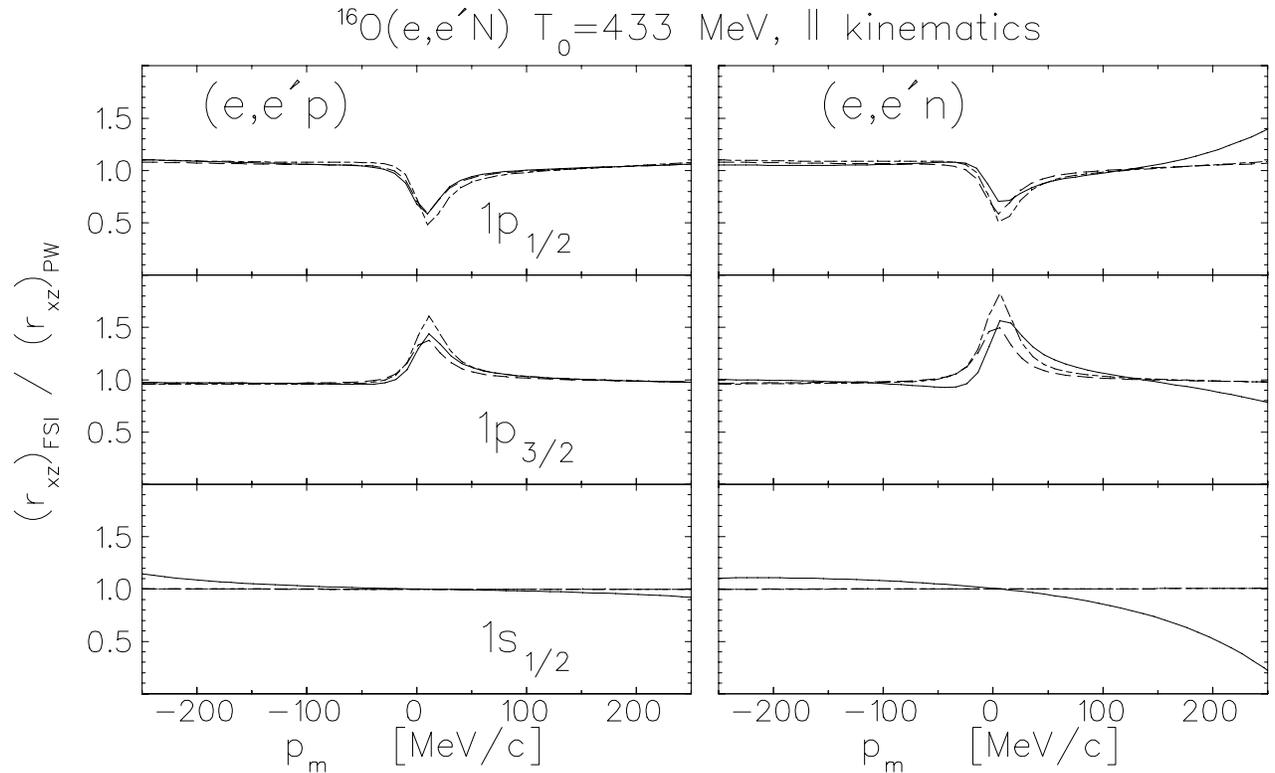,height=4.0in,angle=90} }
\caption{The sensitivity of recoil polarization in parallel kinematics
to FSI is illustrated by comparing $r_{xz}=P^\prime_x / P^\prime_z$ to 
plane-wave (PW) calculations for the $^{16}$O$(\vec{e},e^\prime \vec{N})$ 
reaction at $T_0 = 433$ MeV.
The left (right) column shows proton (neutron) knockout and the three 
rows show calculations for $(1p_{1/2})^{-1}$, $(1p_{3/2})^{-1}$, and 
$(1s_{1/2})^{-1}$ final states.
Dashed and dash-dotted curves represent EEI and EDAD1 optical-model 
calculations, while solid curves include channel coupling for EEI.
}
\label{fig:para_ratio}
\end{figure}

\begin{figure}[htb] 
\centerline{
\strut\psfig{file=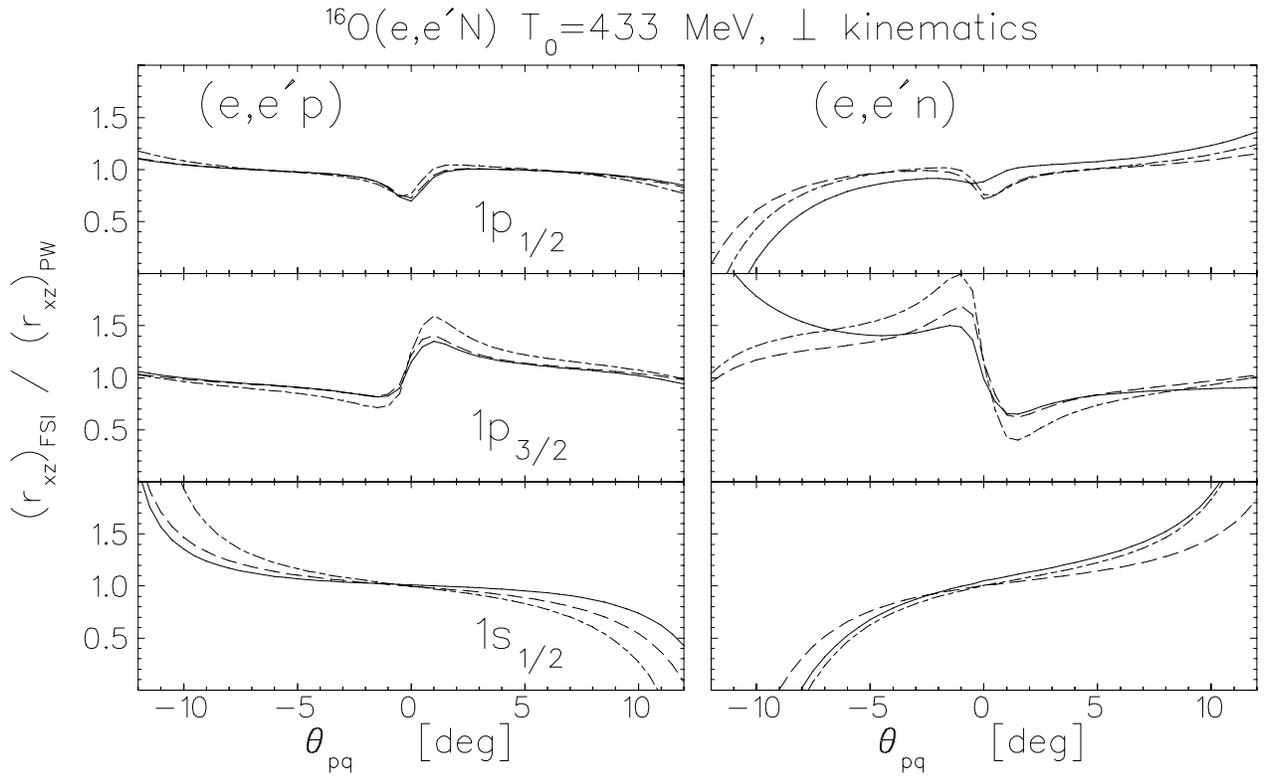,height=4.0in,angle=90} }
\caption{The sensitivity of recoil polarization in quasiperpendicular 
kinematics to FSI is illustrated by comparing 
$r_{xz}=P^\prime_x / P^\prime_z$ to 
plane-wave (PW) calculations for the $^{16}$O$(\vec{e},e^\prime \vec{N})$ 
reaction at $T_0 = 433$ MeV.
See Fig.\ \protect{\ref{fig:para_ratio}} for legend.}
\label{fig:perp_ratio}
\end{figure}

In Figs.\ \ref{fig:para_ratio} and \ref{fig:perp_ratio}
we compare model calculations of the polarization ratio to their plane-wave 
limits for parallel and quasiperpendicular kinematics, respectively.
FSI effects vary relatively slowly with missing momentum for 
$p_m \lesssim 250$ MeV/c, 
but the models quickly diverge from each other thereafter.
Sensitivity to channel coupling in the final state is indicated
by differences between dashed and solid curves, which are both
based upon the EEI model but with the latter including channel coupling. 
Sensitivity to the choice of optical potential is indicated by the
dashed and dash-dotted curves based upon the EEI and EDAD1 models,
respectively.
The EDAD1 potential is a global optical model fitted using 
Dirac phenomenology by Cooper {\it et al}.\ \cite{Cooper93} to proton elastic
scattering data covering a wide range of energy and target mass, and
represents a distinctly different approach than the EEI model.
Also note that the IA2 interaction \cite{Kelly94a} gives results (not shown)
that are practically indistinguishable from EEI over this range of $p_m$.

Figure \ref{fig:para_ratio} show that final-state interactions  
have relatively little effect upon $P^\prime_x/P^\prime_z$ for p-shell 
knockout in parallel kinematics except in the immediate vicinity of the 
node in the momentum distribution where the cross section becomes very 
small anyway.
Not surprisingly, FSI corrections and model dependence are minimized near 
the peaks of the missing momentum distributions for each shell.
Optical model distortion for spin-orbit partners are opposite in direction
and tend to balance for closed shells.
Although s-shell knockout is insensitive to differences between optical models,
the effects of channel coupling are somewhat larger than for the p-shell.
Nevertheless, these effects are minimal for $p_m=0$ and nearly antisymmetric
with respect to the sign of $p_m$ for parallel kinematics.
Therefore, it appears that when $Q^2 \gtrsim 0.5$ (GeV/c)$^2$ 
uncertainties in $r_{xz}$ due to final-state interactions are only at the 
few percent level near the peaks of the missing momentum distributions for 
single-nucleon knockout in parallel kinematics. 

Recoil polarization ratios appear to be more sensitive for quasiperpendicular 
than for parallel kinematics to variations of the FSI model.
The EDAD1 optical potential generally produces larger distortion corrections
to these ratios than do either the EEI or IA2 potentials.
The small differences between dashed and solid curves in 
Fig.\ \ref{fig:perp_ratio} show that channel coupling has very little
effect upon proton knockout from the p-shell for modest opening angles,
but the effects upon p-shell neutron knockout are substantially larger,
especially on the beam side of the momentum transfer.
FSI corrections to $r_{xz}$ for s-shell knockout are relatively small for 
$\theta_{pq} \lesssim 5^\circ$ where the cross section is fairly large,
but become quite substantial for larger angles with small cross sections.
[Note that when $p_m \approx 0$ at $\theta_{pq} = 0^\circ$ for $1p_{1/2}$ 
in quasiperpendicular kinematics with $T_0 = 433$ MeV, then
$p_m \approx 38$ MeV/c for $1s_{1/2}$ is signficantly larger.]
Fortunately, FSI corrections are approximately antisymmetric with respect to 
${\bf q}$ in quasiperpendicular kinematics, 
such that a symmetric acceptance would tend to reduce the net FSI effect
and variations with respect to model.
Furthermore, in parallel kinematics the FSI effects for spin-orbit partners
also tend to compensate.
Thus, the recoil-polarization ratio for inclusive quasifree knockout
from a closed-shell nucleus centered upon ${\bf p}_m=0$ is expected to be 
approximated well by a plane-wave calculation and small residual FSI 
corrections to not depend strongly upon model.
Therefore, it appears that recoil polarization provides a nearly ideal
probe of modifications of the electromagnetic current in nuclei for
which uncertainties due to final-state interactions are relatively small.

Perhaps the simplest modification of the single-nucleon current
would be a variation of nucleon electromagnetic form factors with density.
Using the quark-meson coupling model,
Thomas {\it et al}.\ \cite{Thomas98a} predict that for p-shell proton knockout
from $^{16}$O this ratio will be suppressed by about 10\% relative to the 
free nucleon at $Q^2 = 0.8$ (GeV/c)$^2$.
Similarly, Lu {\it et al}.\ \cite{Lu98c} predict a 12\% suppression of 
$G_{En}$ in $^3$He at $Q^2 = 0.5$ (GeV/c)$^2$.
The present results suggest that final-state interactions will not obscure
these medium modifications of nucleon form factors.
This effect is predicted to increase with $Q^2$ and also becomes sensitive 
at large $Q^2$ to possible variation of the bag constant.
An upcoming experiment \cite{TJNAF93-049} 
measuring recoil polarization for proton knockout
from $^4$He for several $Q^2$ between 0.8 and 4.0 (GeV/c)$^2$ should be
sensitive to such variations of the bag model.
However, two-body currents such as intermediate isobar
excitation, relativistic distortion of nucleon spinors, or off-shell
form factors may also affect the recoil polarization ratio at the several
percent level.
Thus, because two-body currents are expected to affect neutron and proton
knockout somewhat differently, it becomes important to perform measurements 
for both to distinguish between two-body contributions and modifications
of the one-body current.

\section{Summary and conclusions}
\label{sec:conclusions}

We have developed a model of final-state interactions
for $(\vec{e},e^\prime \vec{N})$ reactions in which coupling between
single-nucleon knockout channels is mediated by potentials obtained by 
folding density-dependent nucleon-nucleon effective interactions with 
nuclear transition densities using the local density approximation.
Coupling to more complicated configurations is represented by optical
potentials based upon the same effective interactions.
All couplings within the model space and all components of the 
nucleon-nucleon interaction except tensor exchange are included.
Hence, the model employs a more realistic description of final state 
interactions and can be employed for higher energies than earlier models.
Although the present applications use a one-body current operator and
uncorrelated wave functions,
the model can be extended to include two-body currents and 
ground-state correlations.

To compare our model of charge exchange FSI with earlier approaches, 
we analyzed the $^{12}$C$(\vec{e}, e^\prime \vec{N})$ reaction at 
$T_0 = 70$ MeV using a simple 4-state coupling scheme based upon pure 
$1p_{3/2}$ and $1s_{1/2}$ hole states.
Although van der Steenhoven {\it et al.} \cite{vdSteenhoven87a} predicted
strong charge-exchange contributions to $(e,e^\prime n)$ under these
conditions using the Lane model, we obtain rather small effects for this
model, in agreement with  Giusti and Pacati \cite{Giusti89}.
However, strong charge-exchange contributions to the $(e,e^\prime n)$
cross section are obtained when the 
$t_{\sigma \tau}$ final-state interaction is included.
Similar findings were also obtained by
Jeschonnek {\it et al}.\ \cite{Jeschonnek94}.
We also find that recoil polarization for $(\vec{e},e^\prime \vec{n})$
is quite sensitive to channel coupling, including the helicity-dependent
components, 
while  $(\vec{e},e^\prime \vec{p})$ remains rather insensitive to these 
complications even for these relatively low ejectile energies.

We studied the $^{16}$O$(\vec{e}, e^\prime \vec{N})$ reactions at
$T_0 = 200$ and 433 MeV, 
kinematics appropriate to experiments at MAMI and TJNAF, 
using a 6-state coupling scheme based upon the independent particle model. 
We find that channel coupling has very little effect upon the proton knockout 
cross section for missing momenta $p_m < 300$ MeV/c,
and that the charge-exchange contribution to neutron knockout decreases as
the ejectile energy increases, 
but that channel coupling remains important for neutron knockout 
at $T_0 = 200$ MeV.
For larger $p_m$ channel coupling has important effects upon the cross
sections for both proton and neutron knockout even when $T_0 = 433$ MeV and
these effects depend strongly upon both nuclear structure and kinematics.

Most of the response functions for proton knockout that would remain
finite in the absence of FSI appear to be relatively insensitive to
channel coupling, but response functions for neutron knockout, especially
those which vanish without FSI, are considerably more sensitive to
channel coupling.
Charge exchange mediated by the $t_{\sigma\tau}$ interaction is the
most important coupling mechanism for $^{16}$O$(\vec{e},e^\prime \vec{N})$,
but quadrupole inelastic scattering can be important also for deformed
targets. 

The polarization transfer observables, $P^\prime_S$ and $P^\prime_L$, 
for proton knockout with modest missing momentum appear to be quite 
insensitive to details of the final-state interaction, 
including channel coupling.
Although the corresponding quantities for neutron knockout are affected
at low energies by channel coupling, these effects decrease with energy 
and become relatively small for $T_0 = 433$ MeV.
FSI model dependence is minimized for parallel kinematics near peaks
of the missing momentum distributions for each shell or 
for inclusive quasifree knockout with momentum acceptance that is 
symmetric about ${\bf p}_m=0$.
Furthermore, these quantities appear to be insensitive to ambiguities
in gauge or off-shell properties of the one-body electromagnetic current
operator.
Hence, recoil polarization provides an ideal means for investigating
the electromagnetic current in the nuclear medium.
To the extent that the one-body current is dominant, 
the simple relationship between $P^\prime_S/ P^\prime_L$ and $G_E/G_M$
provides a means for studying possible density dependence of nucleon
electromagnetic form factors.
However, the role of two-body currents at high $Q^2$ has not yet been
investigated and may be important also.
Therefore, it is important to measure separated response functions
which will provide differing sensitivities to these two mechanisms.

The present formalism would permit many technical improvements to be 
implemented in a relatively straightforward manner, 
including correlated wave functions, expanded model spaces, 
improved electron distortion and initial-state coupling, and 
nonlocal final-state interactions.
Extension to relativistic FSI models is also possible.
However, perhaps the most interesting extensions involve the effective
current operator.
In addition to conventional two-body currents, the quark-meson coupling model 
suggests that nucleon electromagnetic form factors are density dependent. 
The implications of density-dependent form factors can be investigated by
applying the local density approximation to the one-body current operator.
The present results suggest that these effects, and those of two-body
currents, can be studied with relatively little uncertainty due to 
final-state interactions using recoil polarization for energetic ejectiles.

\acknowledgements

The author thanks Professor George Rawitscher for valuable discussions of
the coupled-channels formalism.
The support of the U.S. National Science Foundation under grant PHY-9513924 
is gratefully acknowledged.



\clearpage

\clearpage

\end{document}